\documentclass[sn-mathphys-num,iicol,pdflatex]{sn-jnl}

\usepackage{hyperref}
\hypersetup{unicode,colorlinks=true,allcolors=blue} 
\usepackage{graphicx}
\usepackage{dcolumn}
\usepackage{bm}
\usepackage{amssymb}
\usepackage[version=4]{mhchem}
\usepackage{comment}
\usepackage{tabularx}
\usepackage[normalem]{ulem} 
\usepackage[dvipsnames]{xcolor}

\usepackage[utf8]{inputenc}
\usepackage[T1]{fontenc}
\usepackage{mathptmx}
\usepackage{etoolbox}
\usepackage[separate-uncertainty=false]{siunitx} 

\usepackage[nopostdot,nomain,acronym,toc,style=super]{glossaries}

\newacronym{ac}{AC}{alternating current}
\newacronym{acf}{ACF}{autocorrelation function}
\newacronym{afm}{AFM}{atomic force microscopy}
\newacronym{alc}{ALC}{avoided level crossing}
\newacronym{api}{API}{application programming interface}
\newacronym{aps}{aPS}{atactic polystyrene}
\newacronym{ariel}{ARIEL}{Advanced Rare Isotope Laboratory}
\newacronym{arpes}{ARPES}{angle-resolved photoemission spectroscopy}

\newacronym[sort={b-NMR}]{bnmr}{\(\beta\)-NMR}{\(\beta\)-detected nuclear magnetic resonance}
\newacronym[sort={b-NQR}]{bnqr}{\(\beta\)-NQR}{\(\beta\)-detected nuclear quadrupole resonance}
\newacronym{bca}{BCA}{binary collision approximation}
\newacronym{bpp}{BPP}{Bloembergen-Purcell-Pound}
\newacronym{bf4}{BF\(_4\)}{tetra\-fluoro\-borate}
\newacronym{bmim}{BMIM}{1-butyl-3-methyl\-imidazolium}
\newacronym{bmim-tfsa}{BMIM-TFSA}{\glsdesc{bmim} \glsdesc{tfsa}}
\newacronym{bmim-bmsf}{BMIM-BMSF}{1-butyl-3-methyl\-imidazolium bis(tri\-fluoro\-methane sulphonate)\-imide}
\newacronym{bmim-cl}{BMIM-Cl}{1-butyl-3-methyl\-imidazolium chloride}
\newacronym{bmim-pf6}{BMIM-PF\(_6\)}{1-butyl-3-methyl\-imidazolium hexa\-fluoro\-phosphate}
\newacronym{bmsf}{BMSF}{bis(tri\-fluoro\-methane sulphonate)\-imide}
\newacronym{bsc}{BSC}{\ce{Bi2Se3:Ca}}
\newacronym{btm}{BTM}{\ce{Bi2Te3:Mn}}
\newacronym{bts}{BTS}{\ce{Bi2Te2Se}}

\newacronym{camp}{CAMP}{control and monitor program}
\newacronym{ccd}{CCD}{charge-coupled device}
\newacronym{cdw}{CDW}{charge density wave}
\newacronym{cgs}{CGS}{centimetre-gram-second system of units}
\newacronym{ckn}{CKN}{2Ca(NO\(_3\))\(_2\)-3KNO\(_3\)}
\newacronym{cl}{Cl}{chloride}
\newacronym{cmms}{CMMS}{Centre for Molecular and Materials Science}
\newacronym{codata}{CODATA}{Committee on Data for Science and Technology}
\newacronym{cpu}{CPU}{central processing unit}
\newacronym{create}{CREATE}{Collaborative Research and Training Experience Program}
\newacronym{cw}{CW}{continuous wave}
\newacronym{cwt}{CWT}{capillary wave theory}

\newacronym{dac}{DAC}{digital to analog converter}
\newacronym{daq}{DAQ}{data acquisition}
\newacronym{dc}{DC}{direct current}
\newacronym{deme}{DEME}{N,N-\-diethyl-N-\-methyl-N-(2-methoxy\-ethyl)\-ammonium}
\newacronym{dft}{DFT}{density functional theory}
\newacronym{dmm}{DMM}{digital multimeter}
\newacronym{dos}{DOS}{density of states}
\newacronym{dqt}{DQT}{double-quantum transition}

\newacronym{ecnle}{ECNLE}{elastically collective nonlinear Langevin equation}
\newacronym{edm}{EDM}{electric dipole moment}
\newacronym{efg}{EFG}{electric field gradient}
\newacronym{epr}{EPR}{electron paramagnetic resonance}
\newacronym{esr}{EPR}{electron spin resonance}
\newacronym{endor}{ENDOR}{electron nuclear double resonance}
\newacronym{emim-ac}{\ce{EMIM-Ac}}{1-ethyl-3-methyl\-imidazolium acetate}
\newacronym{epics}{EPICS}{Experimental Physics and Industrial Control System}

\newacronym{fft}{FFT}{fast Fourier transform}
\newacronym{fg}{FG}{fluxgate}
\newacronym{fits}{FITS}{flexible image transport system}
\newacronym{fom}{FoM}{figure of merit}
\newacronym{fsa}{FSA}{bis\-(fluoro\-sulfonyl)\-amide}
\newacronym{fwhm}{FWHM}{full width at half maximum}

\newacronym{gga}{GGA}{generalized gradient approximation}
\newacronym{gpl3}{GPL-3.0 License}{GNU General Public License v3.0}
\newacronym{gui}{GUI}{graphical user interface}

\newacronym{hv}{HV}{high-voltage}
\newacronym{hwhm}{HWHM}{half width at half maximum}

\newacronym{il}{IL}{ionic liquid}
\newacronym{is}{IS}{impedance spectroscopy}
\newacronym{isac}{ISAC}{isotope separator and accelerator}
\newacronym{isol}{ISOL}{isotope separation online}
\newacronym{isosim}{IsoSiM}{Isotopes for Science and Medicine}

\newacronym{kww}{KWW}{Kohlrausch-Williams-Watts}

\newacronym{lcao}{LCAO}{linear combination of atomic orbitals}
\newacronym{lda}{LDA}{local density approximation}
\newacronym{leis}{LEIS}{low-energy ion scattering}
\newacronym{lib}{LIB}{lithium-ion battery}
\newacronym{lj}{LJ}{Lennard-Jones potential}
\newacronym{lsat}{LSAT}{\ce{(La,Sr)(Al,Ta)O3}}

\newacronym{mas}{MAS}{magic angle spinning}
\newacronym{mbe}{MBE}{molecular beam epitaxy}
\newacronym{mc}{MC}{Monte Carlo}
\newacronym{md}{MD}{molecular dynamics}
\newacronym{midas}{MIDAS}{Maximum Integrated Data Acquisition System}
\newacronym{mit}{MIT}{metal-insulator transition}
\newacronym{mnr}{MNR}{Meyer-Neldel rule}
\newacronym{mqt}{mqt}{multi-quantum transition}
\newacronym{msb}{MSB}{magnetically shielded box}
\newacronym{msr}{MSR}{magnetically shielded room}
\newacronym{mud}{MUD}{muon data}
\newacronym{mw}{M\textsubscript{w}}{molecular weight}

\newacronym{nbm}{NBM}{neutral beam monitor}
\newacronym{neb}{NEB}{nudged elastic band}
\newacronym{nedm}{nEDM}{neutron electric dipole moment}
\newacronym{nidaq}{NI-DAQ}{National Instruments USB-6281 multifunction IO}
\newacronym{nim}{NIM}{nuclear instrumentation module}
\newacronym{nmr}{NMR}{nuclear magnetic resonance}
\newacronym{no}{NO}{nuclear orientation}
\newacronym{nqr}{NQR}{nuclear quadrupole resonance}
\newacronym{nserc}{NSERC}{Natural Sciences and Engineering Research Council of Canada}

\newacronym{oa}{OA}{optical absorption}
\newacronym{ode}{ODE}{ordinary differential equation}

\newacronym{pac}{PAC}{perturbed angular correlation}
\newacronym{pad}{PAD}{perturbed angular distribution}
\newacronym{peek}{PEEK}{Polyether ether ketone}
\newacronym{peo}{PEO}{poly(ethylene oxide)}
\newacronym{pf6}{PF\(_6\)}{hexa\-fluoro\-phosphate}
\newacronym{pfg}{PFG}{pulsed field gradient}
\newacronym{pld}{PLD}{pulsed laser deposition}
\newacronym{pmt}{PMT}{photo multiplier tube}
\newacronym{ps}{PS}{polystyrene}
\newacronym{psi}{PSI}{the Paul Scherrer Institute}
\newacronym{pvt}{PVT}{polyvinyl toluene}
\newacronym{pypi}{PyPI}{python package index}

\newacronym{qens}{QENS}{quasielastic neutron scattering}
\newacronym{ql}{QL}{quintuple layer}
\newacronym{qo}{QO}{quantum oscillations}
\newacronym{qzfm}{QZFM}{QuSpin Zero Field Magnetometer}

\newacronym{rbs}{RBS}{Rutherford backscattering}
\newacronym[firstplural=radio frequencies (RF)]{rf}{RF}{radio frequency}
\newacronym{rheed}{RHEED}{reflection high-energy electron diffraction}
\newacronym{rib}{RIB}{radioactive ion beam}
\newacronym{rms}{RMS}{root mean squared}
\newacronym{rtil}{RTIL}{room temperature ionic liquid}

\newacronym{sae}{SAE}{spin-alignment echo}
\newacronym{scm}{SCM}{superconducting magnet}
\newacronym{si}{SI}{International System of Units}
\newacronym{simim}{Si-MIm}{1-methyl-3-trimethylsilylmethylimidazolium}
\newacronym{simim-bf4}{[Si-MIm][Bf\(_4\)]}{1-methyl-3-tri\-methyl\-silyl\-methyl\-imidazolium tetra\-fluoro\-borate}
\newacronym{sims}{SIMS}{secondary ion mass spectrometry}
\newacronym{sf}{SF}{shielding factor}
\newacronym{slr}{SLR}{spin-lattice relaxation}
\newacronym{snr}{SNR}{signal-to-noise ratio}
\newacronym{squid}{SQUID}{superconducting quantum interference device}
\newacronym{srim}{SRIM}{Stopping and Range of Ions in Matter}
\newacronym{ssid}{SSID}{solid-state ionic device}
\newacronym{ssr}{SSR}{spin-spin relaxation}
\newacronym{stm}{STM}{scanning tunnelling microscopy}
\newacronym{sts}{STS}{scanning tunnelling spectroscopy}
\newacronym{swr}{SWR}{standing wave ratio}

\newacronym{ti}{TI}{topological insulator}
\newacronym{tfsa}{TFSA}{bis\-(tri\-fluoro\-methane\-sulfonyl)\-amide}
\newacronym{trim}{TRIM}{Transport and Range of Ions in Matter}
\newacronym{tss}{TSS}{topological surface state}
\newacronym{tmd}{TMD}{transition metal dichalcogenide}
\newacronym{tucan}{TUCAN}{TRIUMF Ultracold Advanced Neutron}

\newacronym{uhv}{UHV}{ultra-high vacuum}
\newacronym{ucn}{UCN}{ultracold neutron}

\newacronym{vdw}{vdW}{van der Waals}
\newacronym{vft}{VFT}{Vogel-Fulcher-Tammann}

\newacronym{xrd}{XRD}{x-ray diffraction}
\newacronym{xrr}{XRR}{x-ray reflection}

\newacronym{ybco}{YBCO}{\ce{YBa2Cu3O_{6+x}}}
\newacronym{ysz}{YSZ}{yttria-stabilized zirconia}

\newacronym[sort={muSR}]{musr}{\(\mu\)SR}{muon spin rotation, relaxation, and resonance}
\newacronym{alc-musr}{ALC-\(\mu\)SR}{avoided level crossing muon spin rotation}
\newacronym{le-musr}{LE-\(\mu\)SR}{low energy muon spin rotation}
\newacronym{lf-musr}{LF-\(\mu\)SR}{longitudinal field muon spin rotation}
\newacronym{rf-musr}{RF-\(\mu\)SR}{radio frequency muon spin rotation}
\newacronym{tf-musr}{TF-\(\mu\)SR}{transverse field muon spin rotation}
\newacronym{zf-musr}{ZF-\(\mu\)SR}{zero field muon spin rotation}

\glsdisablehyper

\usepackage[capitalize,noabbrev]{cleveref}
\crefformat{equation}{Equation~#2#1#3}	
\crefrangeformat{equation}{Equations~#3#1#4 to~#5#2#6}
\crefmultiformat{equation}{Equations~#2#1#3}{ and~#2#1#3}{, #2#1#3}{ and~#2#1#3}

\Crefformat{equation}{Equation~#2#1#3}
\Crefrangeformat{equation}{Equations~#3#1#4 to~#5#2#6}
\Crefmultiformat{equation}{Equations~#2#1#3}{ and~#2#1#3}{, #2#1#3}{ and~#2#1#3}

\makeatletter
\DeclareSIUnit\ppm{\text{ppm}}
\DeclareSIUnit\gauss{\text{G}}
\DeclareSIUnit\hours{\text{h}}
\DeclareSIUnit\e{\text{\textit{e}}}
\DeclareSIUnit\torr{\text{Torr}}
\DeclareSIUnit\T{\tesla}
\DeclareSIUnit\mT{\milli\tesla}
\DeclareSIUnit\uT{\micro\tesla}
\DeclareSIUnit\nT{\nano\tesla}
\DeclareSIUnit\pT{\pico\tesla}
\DeclareSIUnit\fT{\femto\tesla}
\DeclareSIUnit\neV{\nano\eV}
\DeclareSIUnit\bar{\text{bar}}

\graphicspath{{Figures}}

\newcommand{\revi}[1]{#1}
\newcommand{\revdi}[1]{}

\newcommand{\rev}[1]{#1}
\newcommand{\revd}[1]{}

\makeatother
\begin{document}

\title[Performance of the TUCAN Magnetically Shielded Room]{Performance of the TUCAN Magnetically Shielded Room}

\author[1]{\fnm{S.}\sur{Ahmed}}
\author[2]{\fnm{B.}\sur{Algohi}}
\author[3]{\fnm{D.}\sur{Anthony}} 
\author[3]{\fnm{P.}\sur{Berard}}
\author[4]{\fnm{L.}\sur{Barr\'on-Palos}}
\author[3]{\fnm{M.}\sur{Boss\'e}} 
\author[3]{\fnm{A.}\sur{Brossard}}
\author[3]{\fnm{J.}\sur{Chak}}
\author[5]{\fnm{R.}\sur{Curtis}} 
\author[3]{\fnm{C.}\sur{Davis}}
\author[1]{\fnm{R.}\sur{de Vries}} 
\author[3]{\fnm{K.}\sur{Dong}}
\author[6]{\fnm{B.}\sur{Dowie}}
\author[3]{\fnm{K.}\sur{Drury}} 
\author[7]{\fnm{P.}\sur{Fierlinger}}
\author[3,5]{\fnm{B.}\sur{Franke}}
\author*[3]{\fnm{D.}\sur{Fujimoto}}\email{dfujimoto@triumf.ca}
\author[8,9]{\fnm{R.}\sur{Fujitani}}
\author[3]{\fnm{P.}\sur{Giampa}}
\author[3]{\fnm{C.}\sur{Gibson}}
\author[10]{\fnm{R.}\sur{Golub}}
\author[11]{\fnm{K.}\sur{Hatanaka}} 
\equalcont{Deceased.}
\author[1]{\fnm{T.}\sur{Hepworth}}
\author[9,11]{\fnm{T.}\sur{Higuchi}}
\author[3]{\fnm{J.}\sur{Hussain}}
\author[1]{\fnm{A.}\sur{Jaison}}
\author[1]{\fnm{M.}\sur{Katotoka}}
\author[12,13]{\fnm{S.}\sur{Kawasaki}}
\author[5]{\fnm{W.}\sur{Klassen}}
\author[5]{\fnm{E.}\sur{Klemets}} 
\author[14]{\fnm{E.}\sur{Korkmaz}}
\author[10]{\fnm{E.}\sur{Korobkina}}
\author[3]{\fnm{F.}\sur{Kuchler}}
\author[1]{\fnm{M.}\sur{Lavvaf}}
\author[3,1]{\fnm{T.}\sur{Lindner}}
\author[3]{\fnm{N.}\sur{Lo}} 
\author[3]{\fnm{J.}\sur{Malcolm}} 
\author[1]{\fnm{R.}\sur{Mammei}}
\author[3]{\fnm{C.}\sur{Marshall}}
\author[1]{\fnm{J. W.}\sur{Martin}}
\author[3,11]{\fnm{R.}\sur{Matsumiya}}
\author[1]{\fnm{M.}\sur{McCrea}}
\author[5]{\fnm{E.}\sur{Miller}}
\author[15]{\fnm{M.}\sur{Miller}}
\author[11,16,12]{\fnm{K.}\sur{Mishima}}
\author[1]{\fnm{T.}\sur{Mohammadi}}
\author[6]{\fnm{N.}\sur{Murby}}
\author[1]{\fnm{S.}\sur{Pankratz}} 
\author[3,17]{\fnm{R.}\sur{Picker}}
\author[11,18]{\fnm{K.}\sur{Qiao}}
\author[1]{\fnm{T.}\sur{Reimer}} 
\author[5]{\fnm{A.}\sur{Sankaran}}
\author[3,19]{\fnm{W.}\sur{Schreyer}}
\author[3]{\fnm{S.}\sur{Sidhu}}
\author[5]{\fnm{L.}\sur{Smith}}
\author[2]{\fnm{S.}\sur{Stargardter}}
\author[3]{\fnm{R.}\sur{Stutters}} 
\author[1]{\fnm{P.}\sur{Switzer}} 
\author[2]{\sur{Tushar}}
\author[6]{\fnm{B.}\sur{van der Veek}}
\author[5,3]{\fnm{S.}\sur{Vanbergen}}
\author[3]{\fnm{W.T.H.}\sur{van Oers}}
\author[6]{\fnm{D.}\sur{Woolger}}
\author[3]{\fnm{N.}\sur{Yazdandoost}}
\author[5]{\fnm{Q.}\sur{Ye}}
\author[2]{\fnm{A.}\sur{Zahra}}
\author[5]{\fnm{M.}\sur{Zhao}} 

\affil[1]{\orgname{The University of Winnipeg}, \orgaddress{\city{Winnipeg}, \state{MB}, \country{Canada}}}
\affil[2]{\orgname{University of Manitoba}, \orgaddress{\city{Winnipeg}, \state{MB}, \country{Canada}}}
\affil[3]{\orgname{TRIUMF}, \orgaddress{\city{Vancouver}, \state{BC}, \country{Canada}}}
\affil[4]{\orgdiv{Instituto de F\'isica}, \orgname{Universidad Nacional Aut\'onoma de M\'exico}, \orgaddress{\city{Mexico City}, \country{Mexico}}}
\affil[5]{\orgname{The University of British Columbia}, \orgaddress{\city{Vancouver}, \state{BC}, \country{Canada}}}
\affil[6]{\orgname{Magnetic Shields Limited}, \orgaddress{\city{Staplehurst}, \state{Tonbridge, Kent}, \country{UK}}}
\affil[7]{\orgdiv{Department of Physics}, \orgname{Technical University Munich}, \orgaddress{\country{Germany}}}
\affil[8]{\orgdiv{Department of Nuclear Engineering}, \orgname{Kyoto University}, \orgaddress{\city{Kyoto}, \country{Japan}}}
\affil[9]{\orgdiv{Institute for Integrated Radiation and Nuclear Science (KURNS)}, \orgname{Kyoto University}, \orgaddress{\city{Osaka}, \country{Japan}}}
\affil[10]{\orgname{North Carolina State University}, \orgaddress{\city{Raleigh}, \state{NC}, \country{USA}}}
\affil[11]{\orgdiv{Research Center for Nuclear Physics (RCNP)}, \orgname{The University of Osaka}, \orgaddress{\city{Osaka}, \country{Japan}}}
\affil[12]{\orgname{High Energy Accelerator Research Organization (KEK)}, \orgaddress{\city{Tsukuba}, \state{Ibaraki}, \country{Japan}}}
\affil[13]{\orgname{The Graduate University for Advanced Studies (Sokendai)}, \orgaddress{\city{Tsukuba}, \state{Ibaraki}, \country{Japan}}}
\affil[14]{\orgname{The University of Northern BC}, \orgaddress{\city{Prince George}, \state{BC}, \country{Canada}}}
\affil[15]{\orgname{McGill University}, \orgaddress{\city{Montreal}, \state{QC}, \country{Canada}}}
\affil[16]{\orgname{Nagoya University}, \orgaddress{\city{Nagoya}, \state{Aichi}, \country{Japan}}}
\affil[17]{\orgname{Simon Fraser University}, \orgaddress{\city{Burnaby}, \state{BC}, \country{Canada}}}
\affil[18]{\orgdiv{Graduate School of Science}, \orgname{The University of Osaka}, \orgaddress{\city{Osaka}, \country{Japan}}}
\affil[19]{\orgdiv{Physics Division}, \orgname{Oak Ridge National Laboratory}, \orgaddress{\city{Oak Ridge}, \state{TN}, \country{USA}}}

\presentaddresstxt{Present addresses}
\presentaddress{T.~Hepworth: Physikalisches Institut, Universität Heidelberg, Germany; Institut Laue–Langevin, Grenoble, France\par F.~Kuchler: Technical University of Munich, Munich, Germany}

\date{\today}

\abstract{
    The \gls{tucan} collaboration has commissioned a large \glsdesc{msr} to be used for measuring the \gls{nedm} to a precision of \SI{e-27}{\e\cm}. The room is composed of five layers of MuMetal and one layer of copper and sits within the $\lesssim\SI{370}{\uT}$ ambient field produced by the TRIUMF cyclotron. Within this environment, the quasi-static shielding factor was measured to be \revd{\num{3.25 \pm 0.02 e4}}\rev{\num{2.95 \pm 0.02 e4}} at \SI{0.01}{\Hz} with an external peak-to-peak perturbation of \SI{2}{\uT}. Without the large ambient cyclotron field, the shielding factor improves to \revd{\num{3.75 \pm 0.04 e4}}\rev{\num{3.40 \pm 0.03 e4}} at the same perturbation amplitude and frequency. After idealization in the cyclotron field, the residual field at the room center was $B = \SI{1.8\pm0.2}{\nT}$ and the vertical first-order gradient across the central \SI{1}{m^3} ($dB_{\mathrm{z}}/dz$) was \revd{$\SI{-279\pm24}{\pT/\m}$}\rev{$\SI{-307\pm26}{\pT/\m}$}. With additional improvement to the idealization, and with active compensation, we expect the room to be adequate for a \SI{e-27}{\e\cm} \gls{nedm} search.
}

\maketitle
\glsresetall
\section{\label{sec:intro}Introduction}

The \gls{tucan} collaboration is a joint Canadian, Japanese, Mexican, and American venture to measure the \gls{nedm} to a precision of \SI{e-27}{\e\cm}. This is an order of magnitude improvement over the current best measurement~\cite{Abel2020}. The primary advantage held by the TUCAN collaboration is the new spallation-driven superfluid helium \gls{ucn} source recently installed at TRIUMF, Canada~\cite{Martin2010,Schreyer2020a,Ahmed2019b,Ahmed2019a,Kawasaki2019}. Simulations of the source, validated against a prototype operated from 2017--2019~\cite{Ahmed2019}, predict that the fully operational source will deliver \num{1.4 e7} polarized \glspl{ucn} at a density of \SI{220}{UCN/\cm^3} in the \gls{nedm} storage cell~\cite{Sidhu2022,Sidhu2023}. Commissioning and characterization campaigns carried out in 2025--2026 have demonstrated performance consistent with these expectations.

\Glspl{ucn}, or neutrons with energies of a few hundred nano-electronvolts, may be contained in magnetic or physical bottles for hundreds of seconds or more. Stored \glspl{ucn} may be used to measure the \gls{nedm} by applying Ramsey's method of separated oscillating fields~\cite{Ramsey1950}. Under Ramsey's method, the \glspl{ucn} undergo coherent spin precession for hundreds of seconds in a magnetic holding field that is both static and uniform. If there is a non-zero \gls{nedm}, the additional application of a large electric field will either increase or decrease the precession frequency, depending on whether the electric field is applied parallel or anti-parallel to the magnetic field. The \gls{nedm}, $d_{\mathrm{n}}$, can then be calculated as:
\begin{equation}\label{eq:edm}
    d_{\mathrm{n}} = \frac{\hbar\left(\omega_{\uparrow\uparrow}-\omega_{\uparrow\downarrow}\right)}{4E},
\end{equation}
where $\omega_{\uparrow\uparrow}$ ($\omega_{\uparrow\downarrow}$) denotes the precession frequency with an aligned (anti-aligned) electric and magnetic field, and $E$ is the electric field magnitude~\cite{Higuchi2024}. 

For precision measurements using this scheme, the magnetic field must be extremely stable over the course of the experiment, otherwise field drifts could mimic a \gls{nedm} signal. While an atomic magnetometer cohabiting the \gls{ucn} storage cell and an array of atomic magnetometers external to the cell~\cite{Martin2015,Klassen2020,Klassen2024} will be used to correct for field fluctuations and gradients, passive magnetic shielding remains a critical component of this measurement. \rev{A quiet magnetic environment is directly linked to the neutron's decoherence time, which determines how long polarized \gls{ucn} may be stored in the chamber, and in turn determines the statistical sensitivity of the measurement~\cite{Sidhu2022,Sidhu2023}. Additionally, magnetic field gradients coupled to the apparent magnetic field observed by \gls{ucn} in motion produces a false \gls{nedm} systematic error which is difficult to eliminate.}

Magnetic shielding is generally constructed of highly permeable metal: since magnetic flux preferentially travels through the metal, external fields are redirected around the contained volume~\cite{Sumner1987}. There are two underlying mechanisms for passive shielding: for static or slowly evolving fields, microscopic ferromagnetic domains in the material are reoriented, redirecting the magnetic flux~\cite{Jiles1984}. If all the domains in the material have been reoriented, then the material is said to be saturated and will not provide any further shielding at higher fields. This limitation may be overcome with the use of concentric shells: the inner shells are only exposed to fields which have already been shielded by the outer shells. The benefit is multiplicative~\cite{Bidinosti2014,Sumner1987}. The distance between shells is important since the DC field continues to decay for a distance after any given shell~\cite{Bidinosti2014,Sumner1987,Yamazaki2006}. At fields evolving faster than \SI{\sim1}{\Hz}, eddy currents are generated in the material and shielding performance becomes dominated by electrical conductivity and connectivity~\cite{Hoburg1995}. A highly conductive layer, typically aluminum or copper is used for this AC shielding. 

Shielding performance may be characterized by the shielding factor and the residual field. The former may be defined as the fractional decrease in the magnitude of a magnetic field fluctuation as a result of the shielding, and the latter is the absolute field within the shielded volume under stable environmental conditions. Since shielding factor varies as a function of both amplitude and frequency, a common standard in the community is to report shielding performance with a peak-to-peak field of $B_{\mathrm{pp}} = \SI{2}{\uT}$, oscillating at \SI{0.01}{\Hz}.

The performance requirements set out by the \gls{tucan} collaboration on the central \SI{1}{\m^3} of the \gls{msr} are as follows: (a) a shielding factor greater than \num{e5} at \SI{0.01}{\Hz}, (b) residual fields with a magnitude less than \SI{1}{\nT}, (c) stability on the \si{\pT} level over minutes, and (d) first order field gradients less than \SI{100}{\pT/\m}~\cite{Mammei2020}. These requirements are set from the \gls{nedm} precision goal of \SI{e-27}{\e\cm}: a shielding factor of \num{e5} would reduce external \SI{100}{\nT} fluctuations to \SI{<1}{\pT}, facilitating requirement (c). A low residual field necessitates small field gradients inside the room. These gradients in turn contribute directly to systematic errors in the \gls{nedm} measurement~\cite{Abel2021}. All of these requirements shall be achieved within the challenging \SI{370}{\uT} vertical magnetic field created by the TRIUMF cyclotron~\cite{Higuchi2022}, and the spatial constraints within the experimental area. The hall restricts the exterior dimensions to \SI{3.5}{\m} on each face, whereas the internal dimensions must exceed a minimum of \SI{1.8}{\m} for the \gls{nedm} experiment. 

As of this writing, the best magnetically shielded large volumes are the 8-layer BMSR-2 at the Physikalisch Technische Bundesanstalt~\cite{Bork1980,Thiel2007} (shielding factor \num{7.5e4}\rev{, later upgraded to \num{3e5} with the addition of active compensation coils~\cite{liu2021}}) and the new 7-layer \gls{msr} at \gls{psi} (shielding factor \num{e5})~\cite{Ayres2021b,Ayres2022}. Both also achieve sub-nanotesla residual fields. To highlight the effect of additional layers, a high-quality 3-layer room built at Harbin, China achieved shielding factors of 5039, 8748, and 5652 for each orthogonal axis, yet with residual fields of \SI{130}{\pT}~\cite{Sun2021}. In this case, fewer layers has resulted a lower dynamic shielding performance (shielding factor), but with careful idealization they achieve \revd{comparably}\rev{extremely} low residual fields. \rev{In the central \si{m^3}, they achieved residual fields $B_z<\SI{100}{\pT}$ and field gradients $\SI{<60}{\pT}$. This is likely the state of the art in terms of static shielding.} Also of note is the PanEDM two-layer \gls{msr} at the Institut Laue-Langevin, with a shielding factor of \num{278}. The PanEDM \gls{msr} also has a three-layer insert for \gls{nedm} measurements, with a shielding factor of 4700~\cite{Altarev2014,Altarev2015b}. Combined, these should produce a room with a shielding factor on the order of \num{e6}.

In this work we present the \gls{tucan} 6-layer \gls{msr} and describe its design, construction, and initial characterization. We will present first results for the shielding factor, the residual field, and the field gradients at the room center. 

\section{\label{sec:design}Design and Construction}

\begin{figure}[ht]
    \centering
    \includegraphics[keepaspectratio=true,width=0.9\columnwidth,trim=0cm 0cm 0cm 0cm, clip=true]{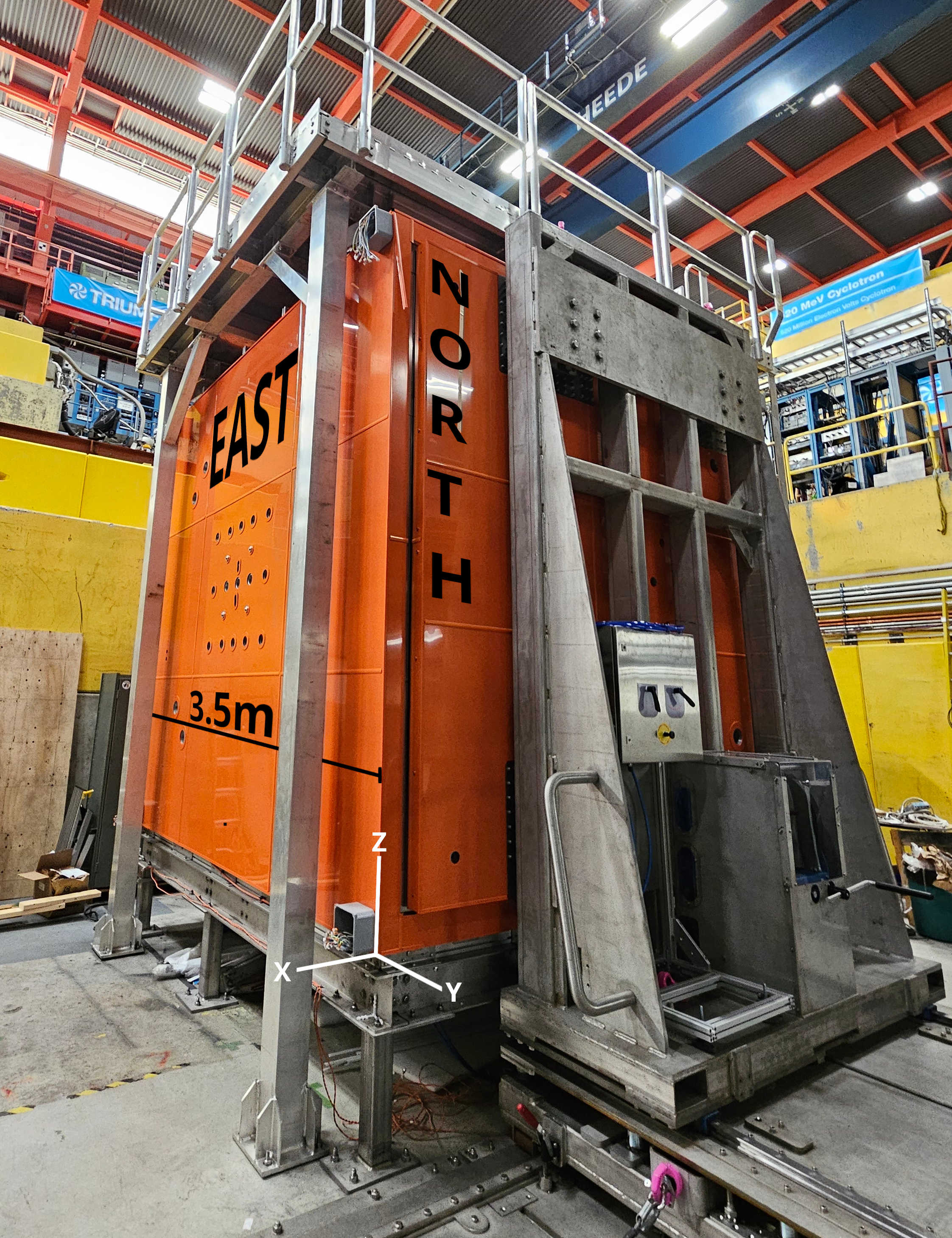}
    \caption{The \glstext{tucan} \glstext{msr} has a door on the North side with two axes of motion. The orange cladding protects the outer MuMetal layer from accidental impact. The aluminum structure built over the \glstext{msr} provides future support for ambient field compensation coils and access to the roof for personnel. Pressure controls for the bladders and pneumatic locks are located on the door support. The door is operated by a pair of manual cranks.}
    \label{fig:msr}
\end{figure}

The \gls{tucan} \gls{msr} is constructed of five layers of annealed high-permeability ASTM A753 Alloy 4 MuMetal (\SI{80}{\percent} Ni, \SI{\sim15}{\percent} Fe, \SI{5}{\percent} Mo) for low-frequency and static magnetic shielding, and one layer of pure copper for high-frequency shielding. The MuMetal alloy saturates at an internal field of \SI{0.75}{T}. Each ferromagnetic layer is individually idealizable and electrically isolated from every other layer (except for a singular grounding point on the South wall), as described further in \cref{sec:res:prep}. \Cref{tab:msrlayers} summarizes the layer dimensions and organization. \revi{The completed room is pictured in \cref{fig:msr}.}

\revi{An aluminum extrusion frame provides the main structure for all layers. For MuMetal layers, the frame is covered by a PVC substrate for electrical isolation and distributed mechanical support, whereas the copper layer is mounted directly to the frame. The MuMetal is retained to the isolating substrate by MuMetal cover strips and identically-sized PVC retaining strips, all of which are fastened by Grade 5 titanium fixings. \Cref{fig:msr2} shows the frame with PVC substrate (white panels), with the fixings covered in electrical tape. The silver of the MuMetal of the preceding layer is visible through the roof framing, as are the dark grey PVC retaining strips. The white styrofoam and pink plastic were temporary to preserve the cleanliness between layers during construction.}

\revi{All MuMetal layers consist of two stacked MuMetal panels to increase the effective layer thickness, with MuMetal cover strips spanning the space between adjacent panels. The cover strips are arranged to fully cover the perimeter of each MuMetal panel in order to support and retain the panel, span air gaps, and create magnetic continuity. All MuMetal panels in each layer are \SI{\le1940}{\mm} in length and \SI{\le590}{\mm} in width, whereas the cover strips are \SI{\le1800}{\mm} in length and \SI{60}{\mm} in width. Panels are spaced \SI{5}{\mm} apart from adjacent panels: this was identified as the minimum spacing achievable due to the stacking of tolerances in both assembly and manufacturing.  Every MuMetal cover strip contains regularly spaced (by \SI{50}{\mm}) clearance holes for fixings.}

\revi{The high-frequency copper shielding layer is comprised of a single layer of panels. Cover strips span the \SI{2}{\mm} air gaps between panels and were cleaned of oxidation immediately prior to installation. The copper panels are \SI{\le940}{\mm} $\times$ \SI{\le940}{\mm}. The panels and gap plates are fixed together using phosphor bronze studs, A4 stainless steel nuts, and A4 stainless steel washers. The spacing on the perimeter panels is \SI{50}{mm}, where the cover strips and panels overlap by \SI{70}{\mm}.} Design, MuMetal panel annealing, and on-site construction of the \gls{msr} was provided by Magnetic Shields Ltd.~\cite{Holmes2022} 

As shown in \cref{fig:msr}, the dual-axis door comprises the entire North face of the room and is supported by a \rev{SAE} 316 stainless steel frame operated by a pair of hand cranks. The door opens by first moving along the longitudinal axis (away from the room, South$\leftrightarrow$North) spanning \SI{80}{\cm} in \num{\sim30} seconds, and subsequently \SI{315}{\cm} in the transverse direction (West$\leftrightarrow$East) over about a minute. This design was dictated by the spatial constraints in the experimental hall. \revi{The dual-axis door, when fully disengaged, provides an access area of \SI{1800}{\mm} $\times$ \SI{1800}{\mm} to the inside of the \gls{msr}.} To ensure good contact with the body of the room, pneumatic locks may be activated to pull the door tight upon closing. \revi{Each shielding layer within the door overlapped the corresponding layer of the \gls{msr} body, ranging from \SI{142}{\mm} in layer 1 to \SI{65}{\mm} in layer 6. The design maximizes the door-body contact area under the constraints of the inter-layer spacing and obstruction by internal equipment (degaussing cables, pneumatics, etc.).} Additionally, a bladder behind the outer edge of door layers 1, 2, 3, and 5 may be inflated to press the MuMetal flat against the MuMetal of the room. The copper layer needed no bladder as a tongue-and-groove connection was used. \revi{The door was constructed within a positional tolerance of \SI{\pm2.5}{\mm} in relation to the room to ensure minimal air gaps between the tongue and groove.}

Access to the room interior is additionally provided by \revdi{88}\revi{98} access ports distributed across the walls, ceiling, and floor. These access ports are necessary for both equipment services and \gls{ucn} ingress and egress. \revi{The access ports combined account for \SI{1.6}{\percent} of the total internal surface area of layer 6.} To reduce gradients in the room, all penetrations through the shielding are mirrored on the opposing face (with some exceptions for the door). Every port in the room is equipped with an aluminum tube penetrating layers~1--4. These tubes are welded to a copper plate that is bolted to the copper layer. This provides structural support to any devices or \gls{ucn} guides entering the room, protects the MuMetal, and prevents objects from falling between the layers (precluding both electrical shorts and permanently magnetized objects between layers). Especially relevant to this work are the two $3\times5$ aligned arrays of ports on the East and West sides of the room (\cref{fig:msr}). These allow for a manual rod-based mapper system to characterize the \gls{msr}'s residual field and internal gradients. 

Construction was started in early 2023, and was finished in August 2024. Assembly started with the \rev{SAE} 316 stainless steel base frame and door support structure, then the floor components of layers 1, 2, and 3. The copper (layer 4) was then fully installed, followed by the fully assembled door. Layer 5 installation was next, then the remaining components of layers 1--3, working outwards. \Cref{fig:msr2} shows the room at the stage where the installation of layers 4 and 5 were fully completed. Each layer of MuMetal is supported with aluminum extrusion and plastic sheets. At the completion of each stage, the magnetic shielding performance of the room was characterized. Following completion of layer 5, the measured shielding factor indicated a reduced performance margin under the large ambient cyclotron field. To provide additional low-frequency attenuation, an additional inner MuMetal layer (layer 6) was installed. Because of this late inclusion, layer 6 does not use the bladder on the door to close contact to the body of the room, relying instead on a set of adjustable nylon bolts. 

The large vertical magnetic background (up to \SI{370}{\uT}) in which the TUCAN \gls{msr} is immersed, is well-approximated as a magnetic dipole and induces a partially-saturating magnetic flux in the outermost layer of the \gls{msr}~\cite{Higuchi2022}. While a planned set of Helmholtz compensation coils will correct for this field, the improvement of the room performance due to these coils will be limited to about \SIrange{15}{30}{\percent}, as will be shown in the next section.

The TUCAN \gls{msr} is approximately aligned with the geographic cardinal points, allowing the definition of the following coordinate system used in this work: $(+x)\rightarrow$~East, $(+y)\rightarrow$~North, and $(+z)\rightarrow$~Up, as shown in \cref{fig:msr}.

\begin{table}[t]
    \centering
    \setlength{\tabcolsep}{3pt}
    \begin{tabular}{@{}lrlcc@{}}
        \textbf{Layer} & \multicolumn{2}{c}{\textbf{Thickness}} & \textbf{Material} & \textbf{Side Length}\\ \hline
        1-outer & \SI{4}{\mm}   & $(2 \times \SI{2.0}{\mm})$    & MuMetal & \SI{3500}{\mm}\\
        2 & \SI{3}{\mm}         & $(2 \times \SI{1.5}{\mm})$    & MuMetal & \SI{3000}{\mm}\\
        3 & \SI{3}{\mm}         & $(2 \times \SI{1.5}{\mm})$    & MuMetal & \SI{2600}{\mm}\\
        4 & $6-\SI{12}{\mm}$    & $(1-2 \times \SI{6.0}{\mm})$  & copper  & \SI{2550}{\mm}\\
        5& \SI{2}{\mm}          & $(2 \times \SI{1.0}{\mm})$    & MuMetal & \SI{2400}{\mm}\\
        6-inner& \SI{2.4}{\mm}  & $(2 \times \SI{1.2}{\mm})$    & MuMetal & \SI{2250}{\mm}
    \end{tabular}
    \caption{The \glstext{tucan} \glstext{msr} is built out of five layers of MuMetal for DC magnetic shielding and one layer of copper for AC magnetic shielding.  Side lengths are outer dimensions. All layers are cubic. The copper layer is formed of panels with large overlaps, such that there are large regions where the thickness is effectively \SI{12}{\mm}.}
    \label{tab:msrlayers}
\end{table}

%

\section{\label{sec:sf} Shielding Factor}

\subsection{Setup}

The shielding factor is a measure of the room's ability to dampen external magnetic fluctuations. It is defined as the ratio of the field fluctuation with and without the shielding present. We measure the shielding factor by placing a magnetometer at the room center and enclose the room in an external coil. Inside the room, we place a reference coil, centered about the magnetometer. Both coils are driven with sinusoidally-varying currents of amplitude $I$ to produce fields of amplitude~$B$. With this setup, the shielding factor can be calculated as:
\begin{equation}\label{eq:sf}
    S = \left(\frac{B_{\mathrm{ext}}^u}{B_{\mathrm{ext}}^s}\right) \left(\frac{B_{\mathrm{ref}}^s}{B_{\mathrm{ref}}^u}\right) \left(\frac{I_{\mathrm{ext}}^s}{I_{\mathrm{ext}}^u}\right) \left(\frac{I_{\mathrm{ref}}^u}{I_{\mathrm{ref}}^s}\right),
\end{equation}
where the superscripts $u$ and $s$ respectively denote the magnetometer as unshielded or shielded  (i.e. before and after \gls{msr} installation).  The subscripts ``ext'' and ``ref'' denote the external and reference coil respectively. Since the reference coil is installed inside the shielding, the field produced by this coil is unaffected by the \gls{msr} construction, allowing the measurement of this field to act as a correction accounting for changes to, or exchanges of, the magnetometer.  
\begin{figure}
    \centering
    \includegraphics[keepaspectratio=true,width=\columnwidth,trim=0cm 0cm 0cm 0cm, clip=true]{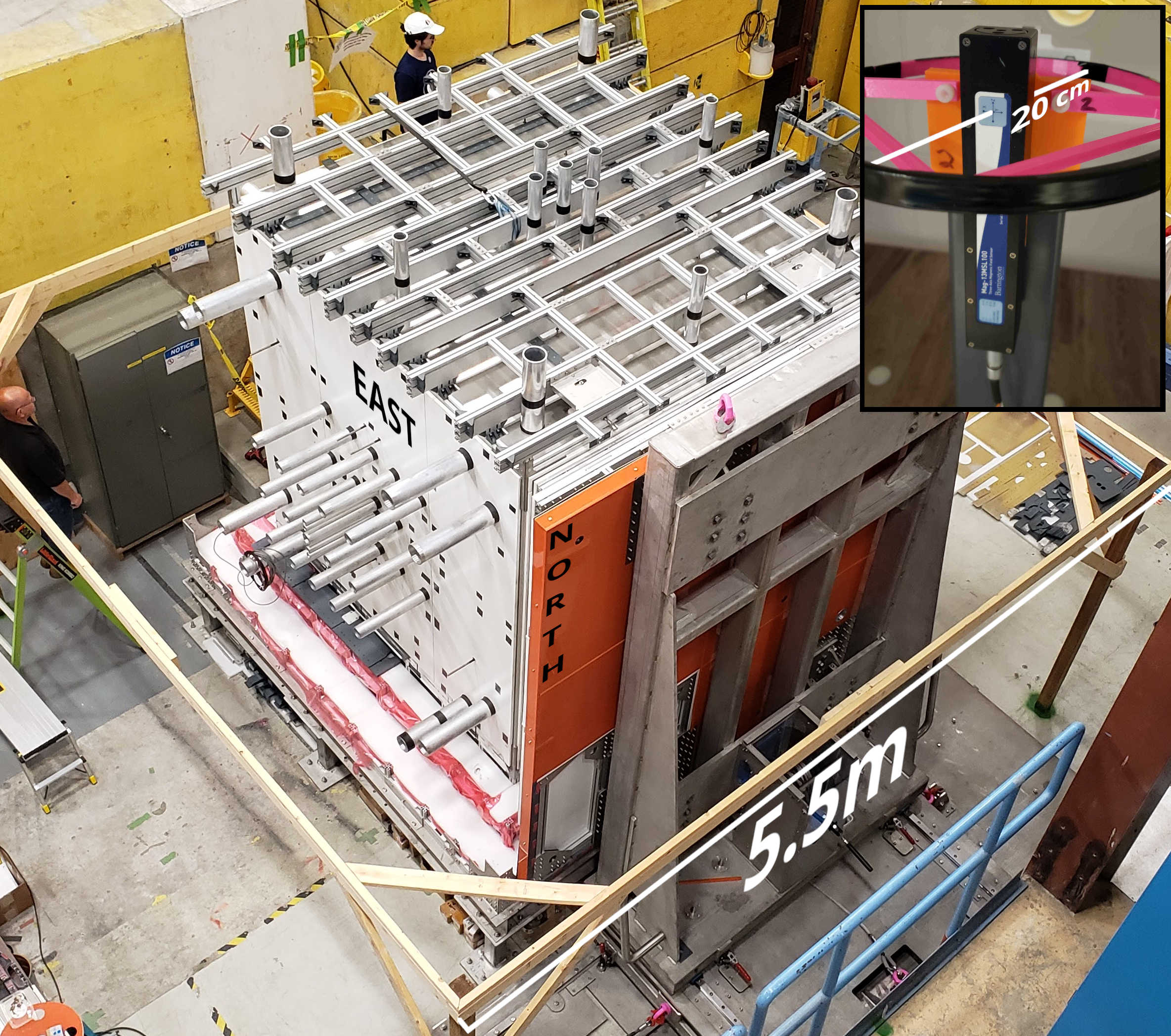}
    \caption{The \glstext{msr} during assembly. By necessity, both floor and door had all layers installed prior to any other work, whereas the rest of the walls were assembled in the order: 4 (copper), 5 (inner), 3, 2, 1 (outermost), then 6 (innermost). The wooden frame supports an external perturbation coil used for the shielding factor measurements. The inset shows the fluxgate and reference coil inside the room.}
    \label{fig:msr2}
\end{figure}

\begin{figure}
    \centering
    \includegraphics[keepaspectratio=true,width=\columnwidth,trim=0.5cm 0.5cm 0.5cm 0cm, clip=true]{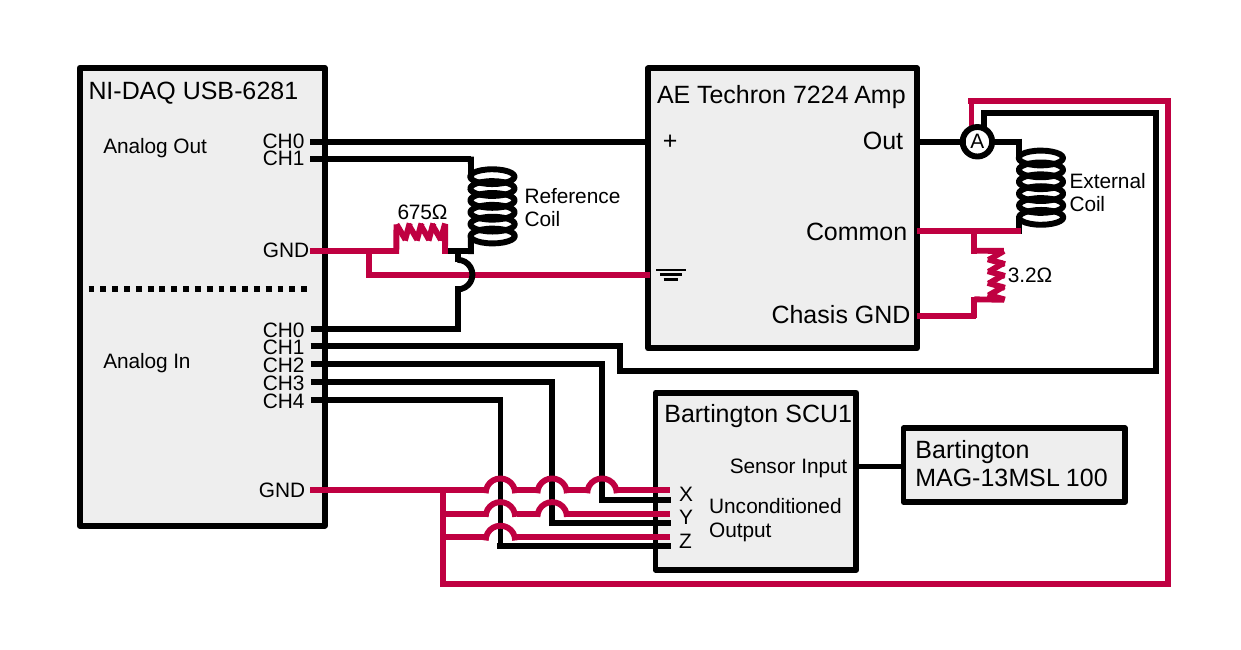}
    \caption{Electronics configuration for supplying current to the external and reference coils, as well as reading out the fluxgate (later replaced by a 3$^{\mathrm{rd}}$ generation three-axis \glstext{qzfm}).}
    \label{fig:sf_diagram}
\end{figure}
%

\begin{figure}[ht]
    \centering
    \includegraphics[keepaspectratio=true,width=\columnwidth,trim=0cm 0cm 0cm 0cm, clip=true]{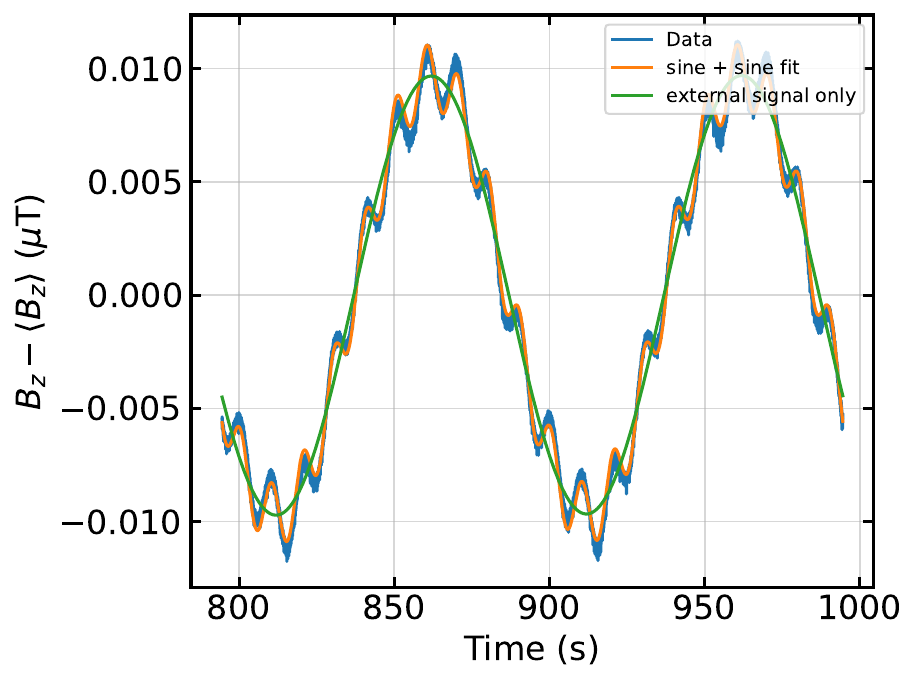}
    \caption{Two cycles of the fluxgate signal at \SI{0.01}{\Hz} with superimposed reference signal at \SI{0.1}{\Hz}, two-sine fit from \cref{eq:fit}, and the component only due to the external field. For all frequencies, at least 10 cycles were measured to ensure that any time-varying background could be corrected properly. The data shown was taken with layers 2 - 5 completed.}
    \label{fig:fit}
\end{figure}

The external field was produced by a 4-turn square coil vertically centered on the internal volume and with side length \SI{5.5}{\m}. This is supported by the wooden frame pictured in \cref{fig:msr2}.  The reference coil was a circular 15-turn, \SI{20}{\cm} diameter coil directly affixed to the magnetometer at the room center by a 3d-printed frame (\cref{fig:msr2} inset). As shown in the electronics configuration drawn in \cref{fig:sf_diagram}, the reference coil was placed in series with a \SI{675}{\ohm} \revd{load}\rev{shunt} to measure the current in the coil. A Pico TA167 current clamp probe was used to measure the current in the large external coil, driven by an AE Techron 7224 power amplifier. A NI-DAQ USB-6281 IO module was used for all data acquisition and amplifier voltage control. Field measurements for the installation of layers 1--5 were taken with a Bartington Mag-13MSL \SI{100}{\uT} fluxgate read out with a Bartington signal conditioning unit (SCU1) \revd{with bandpass} set to the range \SIrange{0}{10}{\kHz}. We applied external fields of peak-to-peak amplitude $B_{\mathrm{ext}}^u = \SI{36}{\uT}$ at the room center since this field is measurable by this fluxgate at shielding factors \num{\sim e5}. The reference signal was chosen to be $\omega_{\mathrm{ref}} = 10\omega_{\mathrm{ext}}$ to ensure that it remained clearly observable and well-separated from the external signal. 

Despite the \SI{1}{\m} gap between the outer wall of the \gls{msr} and the external coil, some field inhomogeneity is present in the oscillating external field. With the \gls{msr} absent, a simple Biot-Savart calculation in the plane of the coil predicts a peak-to-peak field amplitude \SI{\sim83}{\percent} larger at the position of the corners as compared to the center. For all measurements presented in this work, the peak field seen by the MuMetal is well below saturation, and quite small relative to the ambient cyclotron field. 

After installing layer 6, we used a 3$^{\mathrm{rd}}$ generation three-axis \gls{qzfm}~\cite{Shah2018,Zhao_2025} since its lower noise allowed for measurements of the much larger high-frequency shielding factors (\num{\gg e5}). It also permitted measurements with a lower external field of $B_{\mathrm{pp}} = \SI{2}{\uT}$. Notably, the \gls{qzfm} did not use a reference coil due to the \gls{qzfm}'s nominal upper operating frequency of \SI{100}{\Hz}, well under the frequency of the reference coil for a large portion of the measurements (\SIrange{0.1}{1000}{\Hz}). Examination of the reference coil correction factor showed that its exclusion had a relatively negligible effect on the shielding factor determination. Unless specified otherwise, measurements were taken with the cyclotron main magnet energized, as this is the final operating condition of the \gls{nedm} experiment.

At each stage of construction, the shielding factor at a set of 50 frequencies was measured in random order, with the magnetic sensor observing the superposition of the two signals: external and reference. A typical example of the raw data at \SI{0.01}{\Hz} is shown in \cref{fig:fit}. These were modeled as a function of time: 
\begin{equation}\label{eq:fit}
\begin{aligned}
B(t) =~& B_{\mathrm{ext}} \sin(\omega_{\mathrm{ext}} t + \phi_{\mathrm{ext}}) + \\
        & B_{\mathrm{ref}} \sin(\omega_{\mathrm{ref}} t + \phi_{\mathrm{ref}}) + 
        \left(B_0 + B_1t + B_2t^2\right),        
\end{aligned}
\end{equation}
where $\omega_{\mathrm{ext}}$ and $\omega_{\mathrm{ref}}$ are the known frequencies of the external and reference currents, the $\phi_i$ are the phases, and the remaining terms correspond to a phenomenological quadratic background. The amplitudes in \cref{eq:fit} were extracted by least squares fit for all stages of \gls{msr} construction, except when using the \gls{qzfm} (for layer 6), where a two-phase lock-in amplifier analysis (implemented in software) was needed. This phenomenological quadratic background varies on timescales on the order of tens of seconds and is seemingly related to the relaxation of the MuMetal to changes in the magnetic environment -- either from changes in frequency, or from stopping and starting the external coil. To account for this background, at least 10 cycles were taken at each frequency. Across all the completed room data in \cref{fig:sf_compare}, the background parameters were at most $|B_0| < \SI{160}{\pT}$, $|B_1| < \SI{25}{\pT/s}$, and $|B_2| < \SI{1}{\pT/s^2}$. The backgrounds were always determined by least-squares fit, and when using the lock-in amplifier analysis, were determined separately and subtracted before the amplifier code was applied. During fitting, both frequencies were fixed to their known values. 

The two-phase lock-in amplifier was implemented as follows: first, (a) remove any signal background as described above, such that the signal fluctuates about zero. To extract the amplitude at the target frequency, we (b) multiply the signal with a sinusoid at the target frequency. Integrating now zeros all fluctuating components except that at the target frequency, which instead averages to $\frac{1}{2}B_{\mathrm{amp}}\cos\phi$, where $\phi$ is the phase difference, and $B_{\mathrm{amp}}$ is the amplitude of interest. Averaging was accomplished by first (c) applying an 8$^{\mathrm{th}}$ order butterworth lowpass filter with threshold $0.5\omega_{\mathrm{ext}}$, removing any extracted signal at multiples of the target frequency. We then (d) average the resulting time series, omitting the first half which may contain some artifacts from the filtering process. We repeat the process with another sinusoid whose phase is separated by $\frac{\pi}{2}$ from the first. We then (e) add the resulting averages in quadrature, removing the effect of the phase, then (f) multiply by two to recover $B_{\mathrm{amp}}$. The uncertainty due to this method was estimated from the standard deviation of the time series in step (d), which was then propagated through the rest of the analysis.

\revd{Measurement uncertainty estimates reported here are statistical in nature, and are propagated through from the least-squares fit or lock-in amplifier. We} \rev{In addition to statistical errors, we} have also accounted for several sources of systematic errors: using the amplitude of the oscillating signal decouples the measurement from changes in the static environmental fields, the sensor response is tracked via the reference coil, and differences in applied field (reference or external) are corrected by monitoring the current in each coil. Mechanical differences over the course of the year were carefully controlled: the position of the external coil was marked on the concrete floor in paint, the magnetometer stand was bolted to the floor of the room so as to have fixed position, and parts near the sensor were 3D-printed in PLA for mm-accuracy. Systematic errors related to the external coil are small: to lower the shielding factor by \SI{1}{\percent}, the wooden frame would have to be displaced by about \SI{35}{\cm}, much larger than the estimated sub-\si{\cm}-level displacements. Corrections to the shielding factor from the reference coil were also small. 

Because the background fields were too large to operate the \gls{qzfm} used in measuring layer 6, the shielding factor was calculated as $S = cI^u_{\mathrm{ext}}/B^s_{\mathrm{ext}}$, where 
$c = \SI{0.85\pm0.03}{\uT/\A}$ 
was the coil constant of the external coil. This was measured after the start of the \gls{msr} construction, reassembling the coil outside the measurement hall. The coil response was quite linear in the range of \SIrange{14}{19}{\uT}. Variability in this coil constant is one uncorrected systematic error, since in the case of the \gls{qzfm} measurement, the shielding factor is proportional to coil constant. Other than the one reassembly needed for the coil constant measurement, the external coil remained assembled and in-place throughout the room construction. 

\rev{To account for differences in sensor gains, back-to-back measurements of the external signal were taken with both fluxgate and \gls{qzfm}. The current-corrected ratio, averaged over frequencies less than \SI{0.3}{\Hz}, shows that the fluxgate reports a \SI{10.36\pm0.04}{\percent} larger signal than the \gls{qzfm}. This layer 6 measurements have been corrected by this factor, such that the results are self-consistent with the fluxgate readings. }

\subsection{Results}

\begin{figure}
    \centering
    \includegraphics[keepaspectratio=true,width=\columnwidth,trim= 0cm 0cm 0cm 0cm, clip=true]{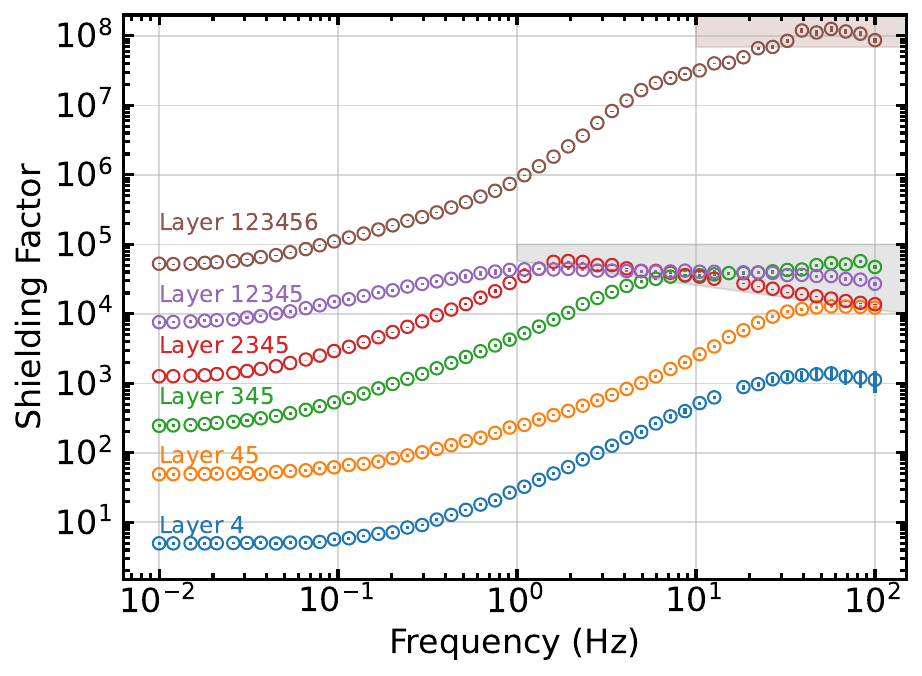}
    \caption{\Glstext{msr} shielding factor (\cref{eq:sf}) at the completion of each layer, measured with external peak-to-peak field of \SI{36}{\uT}. Measurements for all layers were taken with a fluxgate, with the exception of layer 6, where a \gls{qzfm} was used. Shaded areas indicate the approximate noise floors of the two sensor types used.}
    \label{fig:sf_compare}
\end{figure}

Layer-by-layer shielding factors are shown in \cref{fig:sf_compare}. Clearly, the addition of MuMetal layers improves the shielding factor. A plateau is reached at low frequency, known as the quasi-static shielding factor, which may be used to characterize the overall performance of the room. The quasi-static shielding factors at each stage of construction are reported in \cref{tbl:sf_measured}. With an external peak-to-peak amplitude of \SI{36}{\uT}, a final shielding factor of \revd{\num{5.82\pm0.05 e4}}\rev{\num{5.27 \pm 0.04 e4}} was measured at \SI{0.01}{\Hz}. By taking the ratio of the shielding factors at the completion of each layer we can estimate the quasi-static shielding factors of each layer in isolation. In general, additional layers improved the performance by a factor of 5--7. 

\begin{table}
    \begin{tabularx}{\columnwidth}{>{\raggedleft\arraybackslash}X>{\raggedleft\arraybackslash}X>{\raggedleft\arraybackslash}X}
        \textbf{Layers completed} & \textbf{Shielding Factor} & \textbf{Improvement ($S_i/S_{i-1}$)}\\\hline
           4\textcolor{white}{-5-6} & $\num{4.98 \pm 0.22}\times 10^0$ & \\
           4-5\textcolor{white}{-6}  & \num{4.92 \pm 0.21 e1} & \num{9.8\pm0.6}\\
          3-4-5\textcolor{white}{-6}  & \num{2.45 \pm 0.11 e2} & \num{5.0\pm0.3}\\
         2-3-4-5\textcolor{white}{-6}  & \num{1.26 \pm 0.10 e3} & \num{5.2\pm0.2}\\ 
        1-2-3-4-5\textcolor{white}{-6}  & \num{7.59 \pm 0.35 e3} & \num{6.0\pm0.3}\\
        1-2-3-4-5-6               & \rev{\num{5.27 \pm 0.04 e4}} & \rev{\num{6.9\pm0.3}} 
    \end{tabularx}
    \caption{Measured shielding factors at \SI{0.01}{\Hz} and \SI{36}{\uT} external peak-to-peak amplitude. The improvement is the ratio of the shielding factor in that row with the one above, as a measure of the quasi-static shielding factor of any given layer.}
    \label{tbl:sf_measured}
\end{table}

The shielding factor rapidly increases at higher frequencies due to the transition to induced eddy currents as the shielding mechanism~\cite{Hoburg1995}. Above shielding factors of \num{\sim5e4} the signal falls below \SI{300}{\pT}, the noise floor of the fluxgate, leading to a plateau in results for layers 1--5. Switching to a \gls{qzfm}, as was done with layer 6, removes this plateau until its noise floor, around shielding factors of \num{2 e7} (or fields of \SI{1}{\pT}). Notably, the shielding factor denoted by layer 4 corresponds to the combined effect of both the copper layer and the completed sections of the MuMetal layers: the floors of layers 1, 2, and 3; and the fully assembled door. 

During the final stages of construction, the layers were deliberately grounded together via a set of grounding straps, resulting in a sharp decrease in the shielding factor at frequencies greater than \SI{0.15}{\Hz}. It was found that unintended inter-layer connections caused by the titanium floor posts (which are mounted to the copper layer and penetrate layers 5 and 6) created ground loops. These effectively allowed the external grounding straps to pick up and transmit the external oscillating field directly to the inner volume of the room, highlighting the importance of ensuring no layers are electrically connected. Once the floor posts were electrically insulated, performance was restored. 

\begin{figure}
    \includegraphics[keepaspectratio=true,width=\columnwidth,trim= 0cm 0cm 0cm 0cm, clip=true]{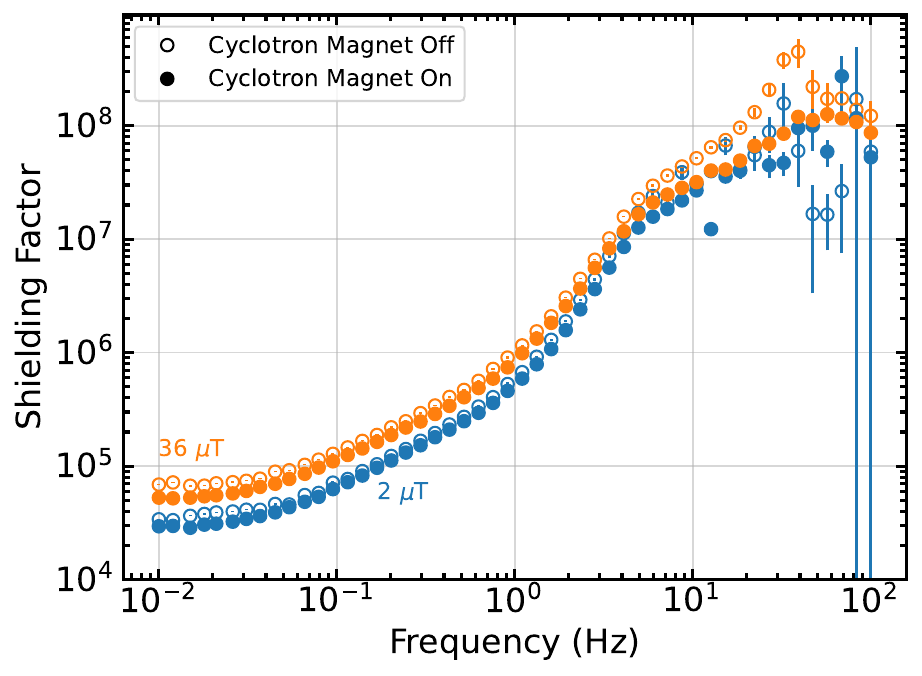}
    \caption{Comparison of the \gls{msr} performance at various external amplitudes and external background fields with all layers installed. The nearby main cyclotron magnet produces a $\lesssim\SI{370}{\uT}$ vertical static field~\cite{Higuchi2022} and is required for \gls{ucn} production, reducing the quasi-static room performance by \SI{15\pm1}{\percent} at \SI{2}{\uT} and \SI{30\pm1}{\percent} at \SI{36}{\uT}. The \gls{qzfm} noise floor is evident for frequencies greater than \SI{\sim7}{\Hz} at \SI{2}{\uT}.}
    \label{fig:compare_cycoff}
\end{figure}

\Cref{fig:compare_cycoff} shows both the effect of reducing the external sinusoidal amplitude and the effect of reducing the static background field, by means of toggling the cyclotron field. The quasi-static shielding factors for each of these four experiments are summarized in \cref{tab:compr}. 

It is well-known that the shielding factor increases with the magnitude of the external fluctuation. From \SI{2}{\uT} to \SI{36}{\uT} peak-to-peak fluctuations the quasi-static shielding factor increases by about \SI{50}{\percent} (\cref{tab:compr}). This difference starts to significantly decrease at frequencies greater than \SI{1}{\Hz}. 

\revd{The effect of the reduced static field is much lower, with 15\% to 30\% improvement to the shielding factor when reducing the ambient field to Earth's field (achieved by turning off the cyclotron magnet). This improvement is reflective of the potential improvement arising from the planned set of coils to compensate for the cyclotron field.}
\rev{An external set of compensation coils will be installed to shim the large static field produced by the cyclotron, with the goal of improving shielding performance. The maximal benefit acheivable by these coils can be estimated by turning off the cyclotron magnet, reducing the ambient field to approximately the Earth's magnetic field. In this case, reflective of perfect cyclotron-compensation, the shielding factor improves by \SIrange{15}{30}{\percent}.}

In comparison to other large \glspl{msr}, the 8-layer BMSR-2 has a quasi-static passive shielding factor of \num{7.5e4}~\cite{Bork1980}. With two fewer layers of MuMetal, the factor of two less for the \gls{tucan} \gls{msr} is an acceptable result. With regards to the room at \gls{psi}, the \gls{tucan} \gls{msr} has one fewer layer of MuMetal and smaller inter-layer spacings. The \gls{psi} room achieved an initial quasi-static shielding factor of \num{e5}~\cite{Ayres2022}. With our \revdi{observed }improvement of $5-7\times$ for each additional layer, extrapolating by one additional layer suggests comparable per-layer shielding performance. However, direct material equivalence to the PSI MSR cannot be inferred due to differences in geometry, inter-layer spacing, and operating environment. While the TUCAN \gls{msr} has an equal number of layers as the PanEDM \gls{msr} (with insert included), the shielding factor of their \gls{msr} can be estimated to be much larger, \num{\sim e6}~\cite{Altarev2014,Altarev2015b}. The PanEDM collaboration attributes the performance of their room to the following design choices: using extraordinarily large panels of MuMetal to minimize joints, thin inner layers, and limiting the ports to few in number and to a small size. 

\begin{table}
    \centering
    \begin{tabularx}{\columnwidth}{>{\centering\arraybackslash}X>{\centering\arraybackslash}X>{\centering\arraybackslash}X}
        \textbf{Peak-To-Peak Field} & \textbf{Cyclotron Magnet} & \textbf{Shielding Factor} \\\hline
        
        \SI{2}{\uT} & Off & \rev{\num{3.40 \pm 0.03 e4}}\\
        \SI{2}{\uT} & On  & \rev{\num{2.95 \pm 0.02 e4}}\\
        \SI{36}{\uT}& Off & \rev{\num{6.88 \pm 0.05 e4}}\\
        \SI{36}{\uT}& On  & \rev{\num{5.27 \pm 0.04 e4}}

    \end{tabularx}
    \caption{Shielding factors at \SI{0.01}{\Hz} for the completed room. The effect of the cyclotron magnet is to decrease the shielding factor by \SI{13\pm1}{\percent} at \SI{2}{\uT} and \SI{23.3\pm0.8}{\percent} at \SI{36}{\uT}. With the cyclotron off, the shielding factor at the higher amplitude increased by \SI{55.9\pm0.5}{\percent}, whereas the improvement was \SI{49.4\pm0.6}{\percent} with the cyclotron on. The value comparable to other rooms is that with the cyclotron off and a \SI{2}{\uT} peak-to-peak amplitude: \revd{\num{3.75 \pm 0.04 e4}}\rev{\num{3.40 \pm 0.03 e4}}.}
    \label{tab:compr}
\end{table}

\section{\label{sec:res} Residual Field}

\subsection{\label{sec:res:prep} Room Preparation}

\Glspl{ucn} moving in a magnetic field with non-zero gradient see a changing field in their rest frame, resulting in a geometric phase shift capable of emulating a \gls{nedm} signal~\cite{Pendlebury2004}.
Therefore, it is imperative that gradients due to remnant magnetization (or any other origin) are limited as much as possible. 

Internal fields and their gradients are minimized by the process of idealization (or demagnetization, if performed in zero external field). Also referred to as degaussing, idealization consists of applying a slowly decaying oscillating field to the ferromagnetic material that composes the walls of the passive shielding~\cite{Thiel2007a,Ayres2024}. The process allows microscopic domains of aligned magnetic moments to more easily align themselves to external fields, thus increasing the amount of flux the material is capable of absorbing, and therefore reducing the internal static field within the room~\cite{Jiles1984}. Idealization is not useful for improving the dynamic shielding performance of the room (\cref{sec:sf}) because the procedure prepares the room only for the external environment present during the idealization procedure. 

\begin{figure}
    \centering
    \includegraphics[keepaspectratio=true,width=0.6\columnwidth,trim=3.5cm 0.5cm 1.5cm 0.75cm, clip=true]{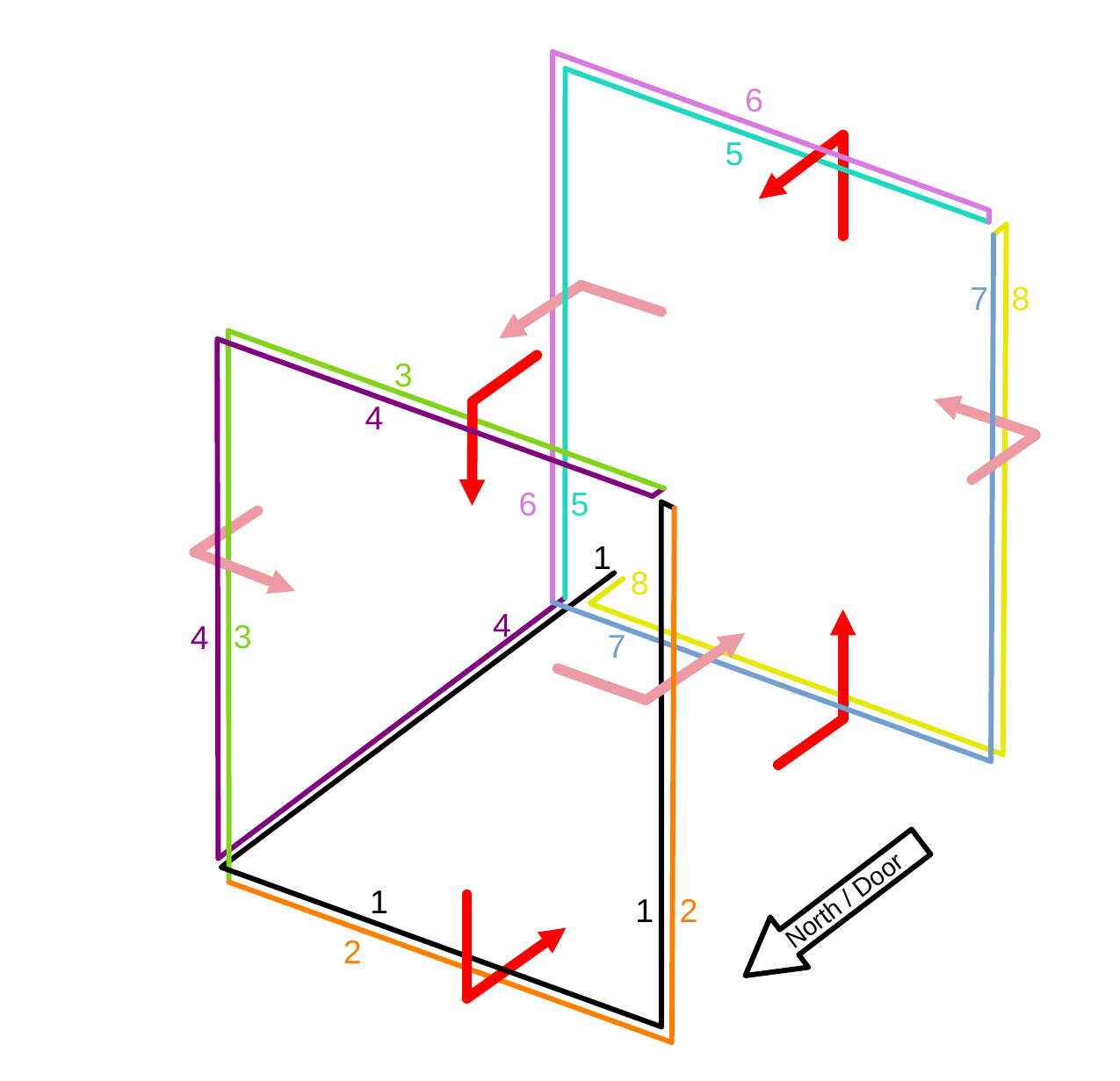}
    \caption{``L'' coils on room edges. Even cables are outside the shielding, odd are inside. For a current flowing along cables in ascending numerical order, fields will be generated along both sets of red arrows. The superposition of these two orthogonal fields result in diagonal flux passing through the walls, allowing for the saturation of all faces at once. This configuration is used on layers 1, 2, and 3. Colors are for legibility only.}
    \label{fig:edge_coil}
\end{figure}

\begin{figure}
    \centering
    \includegraphics[keepaspectratio=true,width=0.6\columnwidth,trim=3cm 2.5cm 1.5cm 0.5cm, clip=true]{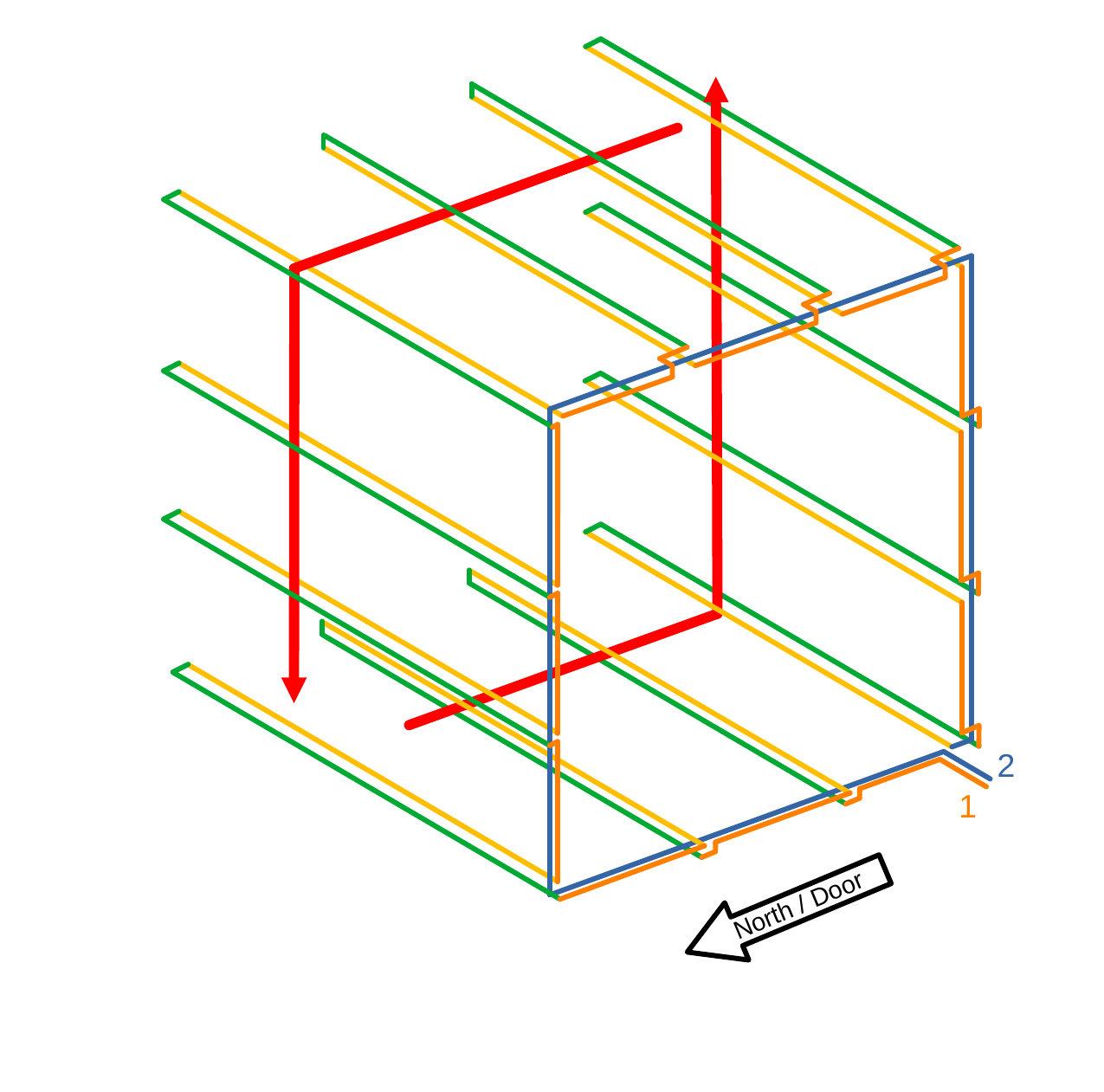}
    \caption{Toroidal coils for one axis, distributed. Green cables are outside the shielding, yellow are inside. Red arrows indicate the direction of the fields if current is supplied to the orange cable, with the return as the blue cable. This configuration, along with an identical orthogonal set, is used in layer 6.}
    \label{fig:dist_vert}
\end{figure}

In the \gls{tucan} \gls{msr}, each ferromagnetic layer has at least one set of idealization coils. The coils can lie on the edges of the cubic \gls{msr} layers (\cref{fig:edge_coil}), or be distributed across the faces (\cref{fig:dist_vert})~\cite{Sun2021}. For either of these layouts, the coils are connected in either a toroidal (\cref{fig:dist_vert}) or ``L'' (\cref{fig:edge_coil}) configuration. In the toroidal configuration, coils about parallel edges (or their interconnecting faces, for the distributed layout) are connected in series, such that the field traces a path around one of the principle axes of the room (e.g. about $\hat{x}$). In the ``L'' configuration, the coils span those edges (faces) used in the toroidal configuration, but are extended into the orthogonal direction. These produce fields which circulate about a tilted axis, allowing the coil set to idealize all walls simultaneously~\cite{Altarev2015,Lavvaf2025}. This has the benefit of shorter idealization procedures, but lacks the simplicity of the toroidal fields. \revi{Coil layouts of the \gls{tucan} \gls{msr} are summarized in \cref{tbl:degaus_coil}.}

\begin{table}
    \begin{tabularx}{\columnwidth}{c>{\raggedleft\arraybackslash}X>{\raggedleft\arraybackslash}X}
        \hspace{0.4cm}\textbf{Layer}\hspace{0.4cm} & \textbf{Primary Coil Set} & \textbf{Secondary Coil Set}\\\hline
        1 & Edge ``L'' (reconfigurable) & \\
        2 & Edge ``L'' & \\
        3 & Edge ``L'' & \\
        5 & Distributed ``L'' & Edge toroidal $\times3$\\ 
        6 & Distributed toroidal $\times2$ & Edge toroidal $\times3$ 
    \end{tabularx}
    \caption{Idealization coil patterns used in the \gls{tucan} \gls{msr}. See the text for a description of these configurations. An example of a set of ``L'' edge coils is shown in \cref{fig:edge_coil}, whereas a set of distributed toroidal coils is shown in \cref{fig:dist_vert}.}
    \label{tbl:degaus_coil}
\end{table}

\begin{figure}
    \centering
    \includegraphics[keepaspectratio=true,width=\columnwidth,trim=0.5cm 0.5cm 0.5cm 0cm, clip=true]{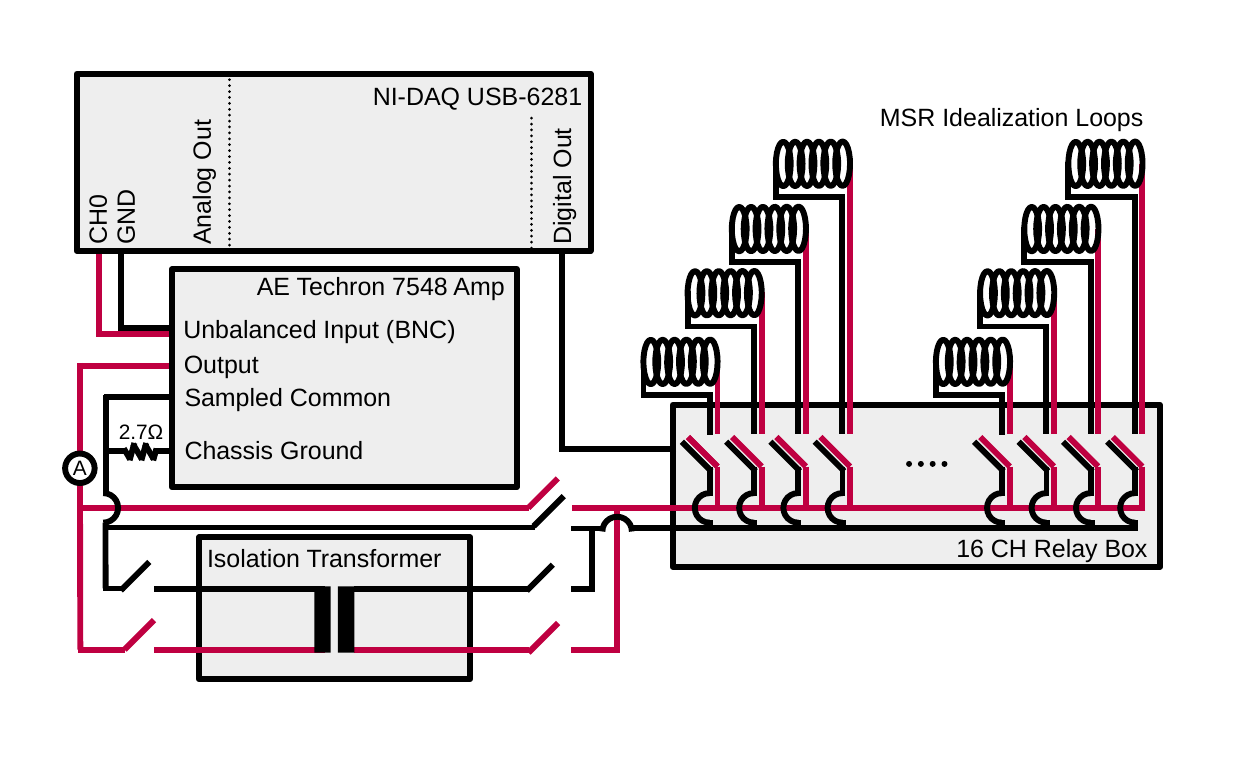}
    \caption{Electronics configuration for supplying current to the idealization loops in the \gls{msr}. The relay box was designed and assembled in-house. Red and black lines denote supply and return.}
    \label{fig:degauss_diagram}
\end{figure}

As shown in \cref{fig:degauss_diagram}, a voltage-controlled AE Techron 7548 Amplifier supplies current to the idealization loops via a set of relay switches designed and assembled at the University of Winnipeg. Both the amplifier and relay box are controlled from a \gls{nidaq}, chosen for its 16-bit analog output channels. The relay box interfaces with the \gls{nidaq} by means of an Arduino, which in turn controls the set of double-pole double-throw mechanical relays. The idealization procedure used in this work starts with the innermost layer 6, then iteratively steps towards the outermost layer 1, then reverses the sequence back to layer 6 again. At each step, we supply a \SI{1}{\Hz} sinusoidal voltage with a time-dependent amplitude to the coils: linearly increasing from zero to max \rev{(\SIrange{10}{20}{\A}, varying by layer)} over \SI{5}{\s}, held at max for \SI{10}{\s}, linearly ramping to zero over \SI{60}{\s}, then held at zero for \SI{5}{\s}. The maximum current was chosen such as to fully saturate each layer. On the second idealization of layer 6, an isolation transformer was introduced into the circuit to remove any current offsets from the final state. To account for this transformer, the frequency was increased to \SI{3}{\Hz} and the duration of the ramp down period increased to \SI{120}{s}. This pattern is roughly modeled after procedures used in similar rooms~\cite{Altarev2015,Ayres2024,Sun2021}.

\subsection{\label{sec:res:meas} Results}

Absolute field values in the \gls{msr} were measured with a \gls{qzfm} fixed to the end of a PVC tube. When the magnetometer is rotated in-place by \SI{180}{\degree}, the sensor sees a reversed field while any measurement offsets are held constant, allowing for an easy calibration of the device. The uncertainty on this calibration is on the order of a few degrees, resulting in $\lesssim\SI{10}{\percent}$ additional error on the field measurements. The tube was inserted through the $3\times5$ array of ports in the West wall of the \gls{msr}, and the field measured with the aforementioned calibration repeated at every position. With the brief idealization procedure described in \cref{sec:res:prep}, and with the $\lesssim\SI{370}{\uT}$ ambient field of the nearby cyclotron, the absolute field in the room is $\lesssim\SI{2}{\nT}$. 

\begin{figure}
    \centering
    \includegraphics[keepaspectratio=true,width=\columnwidth,trim=0cm 0cm 0cm 0cm, clip=true]{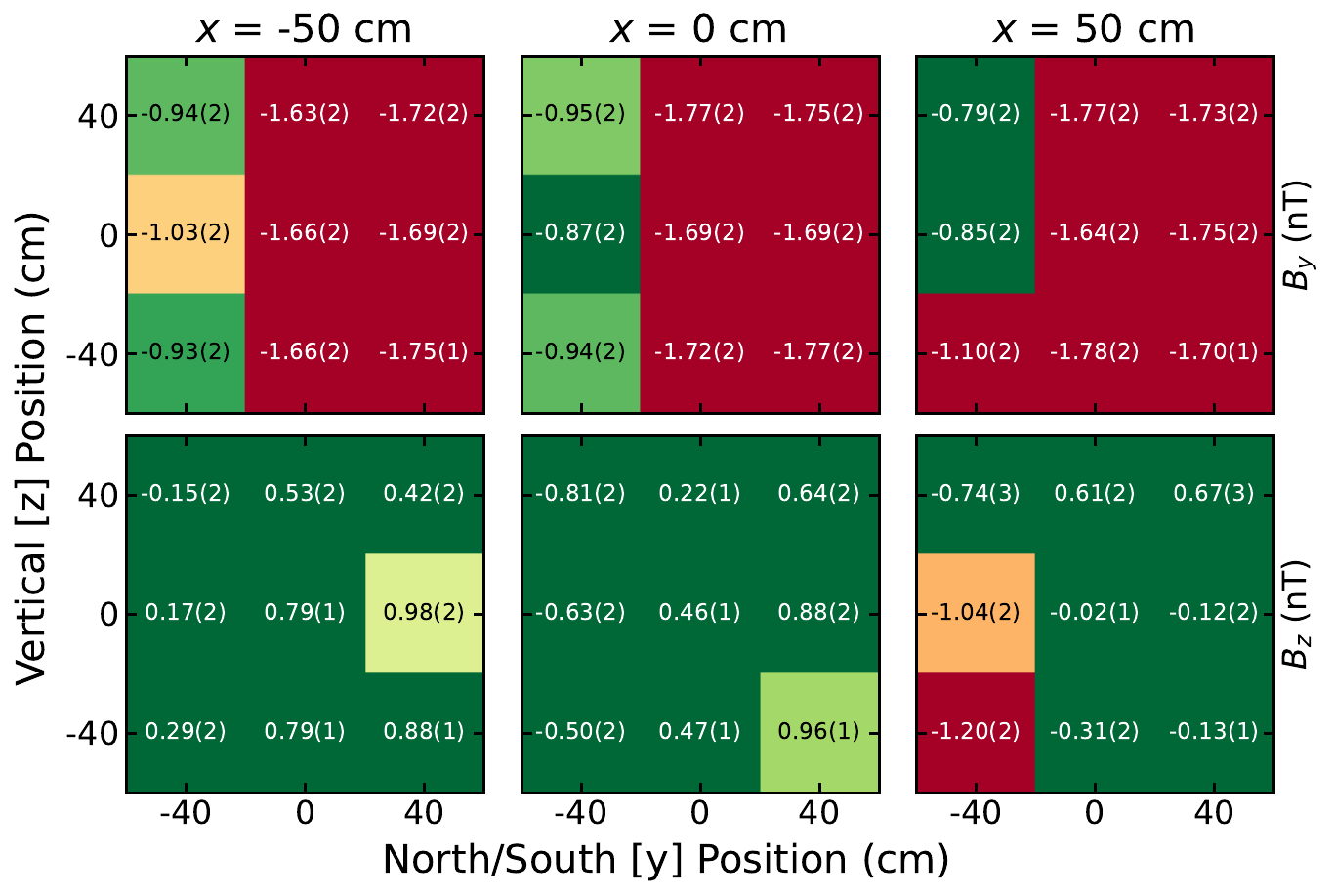}
    \caption{Residual fields in \si{\nT} measured with a \gls{qzfm} and the cyclotron on. The top row is $B_y$, whereas the bottom row is $B_z$. The requirement is \SI{1}{\nT}, green squares satisfy this requirement whereas those which are red do not. While most $B_z$ fields are better than requirement there remains a moderate horizontal field in $B_y$. Axis orientations are as follows: $(+x)\rightarrow$~East, $(+y)\rightarrow$~North, $(+z)\rightarrow$~Up. \Gls{qzfm} internal offsets are accounted for by rotating the device by \SI{180}{\degree} at each position. Due to the axis of rotation, $B_x$ is reported only from a second measurement set through a single penetration in the North door at \revd{$(x, z) = (0,0)$, $B_x(y \in \{0,\SI{-40}{\cm}, \SI{60}{\cm}\}) = \{\num{0.77\pm0.08}, \num{0.69\pm0.07}, \num{0.71\pm0.07}\}~\si{\nT}$}\rev{$(x, z) = (0,0)$, $B_x(y \in \{0,\SI{-40}{\cm}, \SI{60}{\cm}\}) = \{\num{0.76\pm0.02}, \num{0.85\pm0.02}, \num{0.79\pm0.02}\}~\si{\nT}$}.}
    \label{fig:heatmap}
\end{figure}

\Cref{fig:heatmap} shows residual $B_z$ fields which mostly satisfy the requirement of \SI{1}{\nT}, although many $B_y$ values are significantly larger. Brief measurements of $B_x$, which were possible through a singular central hole in the North door, were comparable to those of $B_z$. It is likely the \revdi{large} residual fields measured in the room are due to final few moments of the idealization process. As the oscillating fields \revdi{used in the idealization process }are damped to zero, the MuMetal settles into a state which nominally permits it to absorb as much of the external flux as possible. Any DC offsets (e.g. from the amplifier) or abrupt perturbations (e.g. from disconnecting the circuit) may result in the MuMetal having some non-ideal magnetization. Such magnetization may be exhibited as residual fields and gradients in the room.

\begin{figure}
    \centering
    \includegraphics[keepaspectratio=true,width=0.7\columnwidth,trim=0cm 0cm 0cm 0cm, clip=true]{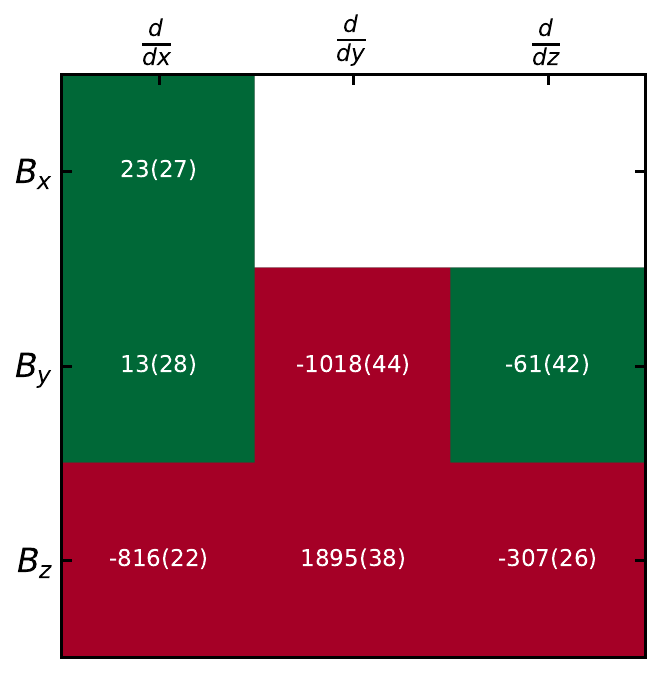}
    \caption{Residual field gradients at the room center as a Jacobian matrix in \si{\pT/\m}, measured with a \gls{qzfm} and the cyclotron on. The requirement is \SI{100}{\pT/\m}. These gradients calculated with a centered differences scheme from the data in \cref{fig:heatmap}. White spaces denote absent measurements, green satisfy the requirement on the field gradient, and red do not.}
    \label{fig:gradient}
\end{figure}

The field gradients at the room center, as found from these initial measurements with the cyclotron field on, are shown in \cref{fig:gradient}. The vertical gradient at the room center, $dB_{\mathrm{z}}/dz$ was \revd{$\SI{-279\pm24}{\pT/\m}$}\rev{$\SI{-307\pm26}{\pT/\m}$}, recalling that the positive direction is up. While this exceeds the requirement on gradients in the \gls{msr} at \SI{100}{\pT/\m} by about a factor of three, this may also be resolved by improving the idealization procedure.

The \gls{tucan} collaboration has several planned strategies to improve these initial results: (a) improve the idealization sequence, (b) installing external compensation coils, and (c) installing an array of internal shim coils~\cite{Bidinosti2014}. \revd{While i}\rev{I}nitial investigations of the idealization procedure have yielded few significant improvements\revd{,}\rev{. However, }we are confident that additional improvements can yet be found, based on the experience of other collaborations~\cite{Altarev2014,Altarev2015,Ayres2024}. Ambient magnetic field compensation coils were planned early in the \gls{msr} planning process to counteract the large static cyclotron field. \Cref{fig:msr} shows a large aluminum frame encasing the \gls{msr}. This is the support frame for a pair of large Helmholtz compensation coils. Finally, an array of 54 internal shim coils is under development, capable of correcting several nanotesla of internal field, as well as reducing gradients. Other techniques, such as magnetic shaking~\cite{Allmendinger2023}, or dynamic active compensation are possible future upgrades. Further studies of the stability and optimization of the residual fields will be included in a future work. 

%

\section{\label{sec:sum} Summary}

The \gls{tucan} collaboration has commissioned an \gls{msr} with five layers of MuMetal and one layer of copper for a \gls{nedm} measurement. The room is cubic with an outer side length of \SI{3.5}{\m} and an inner side length \SI{2.25}{\m}. At the conclusion of the installation of each layer, the dynamic shielding factor was measured as a function of frequency. Generally, the shielding factor increased by a factor of $5-7$ with each layer. Within the large ambient static field generated by the nearby cyclotron, we report a final quasi-static shielding factor of \revd{\num{3.25 \pm 0.02 e4}}\rev{\num{2.95 \pm 0.02 e4}} at \SI{0.01}{\Hz} and an external peak-to-peak amplitude of \SI{2}{\uT}. Without this unique challenge, the quasi-static shielding factor improves to \revd{\num{3.75 \pm 0.04 e4}}\rev{\num{3.40 \pm 0.03 e4}}. While the requirements for an \num{e-27} \gls{nedm} experiment assume ambient field fluctuations on the order of \SI{100}{\nT}, with the required shielding factor of \num{e5} set to reduce these fluctuations to \SI{\sim1}{\pT}, the fluctuations outside and inside the room are yet unmeasured. It may be that the room performance is more than sufficient for the TUCAN \gls{nedm} measurement. 

With a relatively untuned degaussing procedure, and with the cyclotron magnet energized, the absolute residual static field at the room center was $B = \SI{1.8\pm0.2}{\nT}$;
and the vertical gradient at the room center, $dB_{\mathrm{z}}/dz$, was \revd{$\SI{-279\pm24}{\pT/\m}$}\rev{$\SI{-307\pm26}{\pT/\m}$}. These results and their corresponding requirements are summarized in \cref{tbl:summary}. Additional improvements to the residual fields and gradients are planned, including further optimization of the degaussing procedure, installation of compensation systems, and implementation of internal shimming. These efforts are expected to bring the magnetically shielded room to the performance required for a \SI{e-27}{\e\cm} \gls{nedm} search.

\begin{table}
    \centering
    \begin{tabular}{l|l|r}
                            & Requirement       & Measured\\\hline
        Shielding Factor    & \num{>e5}         & \rev{\num{2.95 \pm 0.02 e4}} \\
        Residual Field      & \SI{<1}{\nT}      & $B_{\mathrm{max}} = \SI{1.8\pm0.2}{\nT}$\\
        Field Stability     & \SI{<1}{\pT/min}  & Unmeasured\\
        Field Gradients     & \SI{<100}{\pT/\m} & \rev{$dB_{\mathrm{z}}/dz = \SI{-307\pm26}{\pT/\m}$}
    \end{tabular}
    \caption{The performance requirements on the central \SI{1}{\m^3} of the \gls{msr} as compared to measurements at the room center in this work. These values were measured with the nearby cyclotron main magnet on, which corresponds to a static $\lesssim\SI{370}{\uT}$ ambient vertical field.}
    \label{tbl:summary}
\end{table}

\backmatter

\bmhead{Acknowledgments}

The authors would like to acknowledge the TRIUMF machine shop for their timely work, the Meson Hall support team (M.~Dalla~Valle, A.~Sarkar, P.~Hall-Patch), Beamlines group (T.~Hessels), in addition to the MSL installation team (P.~Harden, S.~Dineen, R.~Willoughby, B.~Dyer, and G.~Spackman) and project engineer (D.~Holmes). Additionally we acknowledge A.~Sher for his simulation work and CMC Microsystems for their provision of CAD software.

\section*{Declarations}

\bmhead{Funding}

Funding for this work and the TUCAN collaboration was supported by the Canada Foundation for Innovation, Innovation Fund grant 36322 and New Initiatives Fund grant 19280; the Canada Research Chair program; the Natural Sciences and Engineering Research Council of Canada (NSERC) SAPPJ-2016-00024, SAPPJ-2019-00031 and SAPPJ-2023-00029; Japan Society for the Promotion of Science (JSPS) KAKENHI grants 18H05230 and 20KK0069; JSPS KENHI grants 17K14307, 19K23442, 20K14487, 21K13940, 22H01236; RCNP CORENet; JST FOREST Program (JPMJFR2237); \rev{the Yamada Science Foundation; the Murata Science Foundation; Yazaki Memorial Foundation for Science and Technology;} the Universidad Nacional Aut\'onoma de M\'exico DGAPA program PASPA; and grant PAPIIT AG102023.

\bmhead{Conflicts of interest}

B.~Dowie, N.~Murby, B.~van~der~Veek, and D.~Woolger are employees of Magnetics Shields Ltd., the firm responsible for the engineering and construction of the room. P.~Fierlinger was a paid consultant for the design of the room. All other authors have no conflicts of interest to disclose. 

\bmhead{Data availability}

The data that support the findings of this study are available from the corresponding author upon reasonable request.

\bmhead{Code availability}

The code used in the analysis of these results are available from the corresponding author upon reasonable request.

\bmhead{Author contributions}

\textbf{S.~Ahmed:} conceptualization (minor support); investigation (minor support); methodology (lead); resources (lead); software (lead); validation (minor support); writing – review \& editing (minor support). \textbf{B.~Algohi:} writing – review \& editing (minor support). \textbf{D.~Anthony:} writing – review \& editing (minor support). \textbf{P.~Berard:} investigation (minor support); writing – review \& editing (minor support). \textbf{L.~Barr\'on-Palos:} investigation (minor support); writing – review \& editing (minor support). \textbf{M.~Boss\'e:} writing – review \& editing (minor support). \textbf{A.~Brossard:} resources (minor support); writing – review \& editing (minor support). \textbf{J.~Chak:} conceptualization (major support); methodology (lead); project administration (major support); validation (minor support); writing – review \& editing (minor support). \textbf{R.~Curtis:} investigation (minor support); writing – review \& editing (minor support). \textbf{C.~Davis:} conceptualization (minor support); funding acquisition (major support); writing – review \& editing (minor support). \textbf{R.~de~Vries:} writing – review \& editing (minor support). \textbf{K.~Dong:} writing – review \& editing (minor support). \textbf{B.~Dowie:} methodology (lead); writing – review \& editing (\revi{major}\revdi{minor} support). \textbf{K.~Drury:} writing – review \& editing (minor support). \textbf{P.~Fierlinger:} conceptualization (major support); formal analysis (minor support); investigation (major support); validation (major support); writing – review \& editing (minor support). \textbf{B.~Franke:} conceptualization (lead); funding acquisition (lead); project administration (lead); resources (major support); supervision (lead); writing – review \& editing (major support). \textbf{D.~Fujimoto:} data curation (lead); formal analysis (lead); investigation (lead); methodology (lead); project administration (major support); resources (lead); software (lead); supervision (lead); validation (lead); visualization (lead); writing – original draft (lead); writing – review \& editing (lead). \textbf{R.~Fujitani:} writing – review \& editing (minor support). \textbf{P.~Giampa:} writing – review \& editing (minor support). \textbf{C.~Gibson:} project administration (lead); writing – review \& editing (minor support). \textbf{R.~Golub:} funding acquisition (minor support); writing – review \& editing (minor support). \textbf{K.~Hatanaka:} conceptualization (lead); funding acquisition (major support). \textbf{T.~Hepworth:} investigation (major support); writing – review \& editing (minor support). \textbf{T.~Higuchi:} conceptualization (lead); funding acquisition (minor support); investigation (major support); writing – review \& editing (minor support). \textbf{J.~Hussain:} writing – review \& editing (minor support). \textbf{A.~Jaison:} investigation (minor support); writing – review \& editing (minor support). \textbf{M.~Katotoka:} investigation (minor support); methodology (major support); resources (major support); writing – review \& editing (minor support). \textbf{S.~Kawasaki:} funding acquisition (major support); writing – review \& editing (minor support). \textbf{W.~Klassen:} conceptualization (major support); investigation (major support); methodology (major support); resources (major support); supervision (major support); validation (major support); writing – review \& editing (minor support). \textbf{E.~Klemets:} investigation (major support); writing – review \& editing (minor support). \textbf{E.~Korkmaz:} funding acquisition (major support); writing – review \& editing (minor support). \textbf{E.~Korobkina:} funding acquisition (major support); writing – review \& editing (minor support). \textbf{F.~Kuchler:} conceptualization (minor support); writing – review \& editing (minor support). \textbf{M.~Lavvaf:} data curation (minor support); formal analysis (major support); investigation (major support); methodology (major support); project administration (minor support); resources (major support); software (major support); supervision (minor support); validation (major support); visualization (major support); writing – review \& editing (minor support). \textbf{T.~Lindner:} funding acquisition (lead); writing – review \& editing (minor support). \textbf{N.~Lo:} writing – review \& editing (minor support). \textbf{J.~Malcolm:} writing – review \& editing (minor support). \textbf{R.~Mammei:} conceptualization (lead); funding acquisition (lead); investigation (lead); methodology (major support); project administration (lead); resources (lead); supervision (major support); writing – review \& editing (minor support). \textbf{C.~Marshall:} conceptualization (lead); methodology (lead); writing – review \& editing (minor support). \textbf{J.~Martin:} conceptualization (lead); funding acquisition (lead); investigation (lead); methodology (lead); project administration (lead); resources (lead); supervision (major support); writing – review \& editing (major support). \textbf{R.~Matsumiya:} writing – review \& editing (minor support). \textbf{M.~McCrea:} conceptualization (major support); methodology (lead); writing – review \& editing (minor support). \textbf{E.~Miller:} writing – review \& editing (minor support). \textbf{M.~Miller:} writing – review \& editing (minor support). \textbf{K.~Mishima:} funding acquisition (major support); writing – review \& editing (minor support). \textbf{T.~Mohammadi:} investigation (minor support); writing – review \& editing (minor support). \textbf{N.~Murby:} project administration (lead); supervision (lead); validation (minor support); writing – review \& editing (minor support). \textbf{S.~Pankratz:} writing – review \& editing (minor support). \textbf{R.~Picker:} conceptualization (major support); funding acquisition (lead); writing – review \& editing (major support). \textbf{K.~Qiao:} writing – review \& editing (minor support). \textbf{T.~Reimer:} writing – review \& editing (minor support). \textbf{A.~Sankaran:} investigation (major support); validation (minor support); writing – review \& editing (minor support). \textbf{W.~Schreyer:} conceptualization (minor support); writing – review \& editing (major support). \textbf{S.~Sidhu:} writing – review \& editing (minor support). \textbf{L.~Smith:} investigation (minor support); writing – review \& editing (minor support). \textbf{S.~Stargardter:} writing – review \& editing (minor support). \textbf{R.~Stutters:} writing – review \& editing (minor support). \textbf{P.~Switzer:} writing – review \& editing (minor support). \textbf{Tushar:} writing – review \& editing (minor support). \textbf{B.~van~der~Veek:} project administration (lead); supervision (lead); writing – review \& editing (minor support). \textbf{S.~Vanbergen:} conceptualization (minor support); resources (minor support); writing – review \& editing (minor support). \textbf{W.T.H.~van~Oers:} conceptualization (major support); funding acquisition (major support); writing – review \& editing (minor support). \textbf{D.~Woolger:} project administration (lead); supervision (lead); writing – review \& editing (minor support). \textbf{N.~Yazdandoost:} writing – review \& editing (major support). \textbf{Q.~Ye:} writing – review \& editing (minor support). \textbf{A.~Zahra:} writing – review \& editing (minor support). \textbf{M.~Zhao:} investigation (minor support); software (major support); validation (major support); writing – review \& editing (minor support).

\bibliography{bib,library}

\end{document}